\documentclass[a4paper,12pt]{article}
\makeatletter

\usepackage[english]{babel}
\usepackage{lmodern}
\usepackage{microtype}

\usepackage{mathtools}
\usepackage{fullpage, euscript, amsfonts, amssymb, amsmath}
\usepackage{diffcoeff}
\usepackage[cal=boondox]{mathalfa}

\usepackage{setspace}
\usepackage[paper=a4paper, 
            inner=22mm,
            outer=22mm,
            top=22mm,
            bottom=30mm,
            headheight=6mm,
            includehead,
            includefoot, 
            heightrounded, 
            twoside]{geometry}
            
\usepackage{pdflscape}

\usepackage{graphicx}
\usepackage{float}

\usepackage{tabularx}
\usepackage{booktabs}
\usepackage{threeparttable}

\usepackage[
    backend=biber,
    style=authoryear,
    sorting=nyt,
    url=false,
    dashed=false]{biblatex}
\usepackage{csquotes}
\DeclareFieldFormat[techreport, report, misc]{title}{\mkbibquote{#1}}
\DeclareFieldFormat[techreport, report]{type}{\mkbibemph{#1}}
\DeclareFieldFormat{eprint}{\texttt{#1}}

\DeclareNameAlias{sortname}{family-given}

\AtBeginBibliography{\defcounter{maxnames}{99}}

\DeclareBibliographyDriver{*}{%
  \usebibmacro{bibindex}%
  \usebibmacro{begentry}%
  \printnames{author}%
  \setunit{\addcomma\space}%
  \printfield{year}%
  \setunit{\addcomma\space}%
  \printfield{title}%
  \newunit\newblock
  \printfield{journaltitle}%
  \setunit{\addcomma\space}%
  \printfield{volume}%
  \setunit{\addcomma\space}%
  \printfield{number}%
  \setunit{\addcomma\space}%
  \printfield{pages}%
  \newunit\newblock
  \printfield{institution}%
  \setunit{\addcomma\space}%
  \printfield{type}%
  \setunit{\addcomma\space}%
  \printfield{number}%
  \usebibmacro{finentry}%
}

\DeclareBibliographyDriver{article}{%
  \usebibmacro{bibindex}%
  \usebibmacro{begentry}%
  \printnames{author}%
  \setunit{\addspace}%
  \printtext[parens]{\printfield{year}}%
  \setunit{\addperiod\space}%
  \printfield{title}%
  \newunit\newblock
  \printfield{journaltitle}%
  \setunit{\addcomma\space}%
  \printfield{volume}%
  \setunit{\addcomma\space}%
  \printfield{number}%
  \setunit{\addcomma\space}%
  \printfield{pages}%
  \usebibmacro{finentry}%
}

\usepackage{multirow}
\usepackage{xcolor}
\definecolor{my_blue}{HTML}{1f77b4}

\usepackage{appendix}

\usepackage{bigfoot}

\usepackage[
    pdftitle={},
    pdfauthor={Maria Elena Filippin},
    hidelinks,
    hyperfootnotes=false
]{hyperref}
\hypersetup{
  colorlinks,
  citecolor=black,
  linkcolor=black}

\newcommand\blfootnote[1]{%
  \begingroup
  \renewcommand\thefootnote{}\footnote{#1}%
  \addtocounter{footnote}{-1}%
  \endgroup
}

\makeatother
\title{Central Bank Digital Currency with Collateral-constrained Banks}

\DeclareNewFootnote{AAffil}[arabic]
\DeclareNewFootnote{ANote}[fnsymbol]

\usepackage{etoolbox}
\makeatletter
\patchcmd\maketitle{\def\@makefnmark{\rlap{\@textsuperscript{\normalfont\@thefnmark}}}}{}{}{}

\makeatletter

\def\thanksAAffil#1{
  \footnotemarkAAffil\protected@xdef\@thanks{\@thanks%
        \protect\footnotetextAAffil[\the \c@footnoteAAffil]{#1}}%
}

\def\thanksANote#1{%
  \footnotemarkANote%
  \protected@xdef\@thanks{\@thanks%
        \protect\footnotetextANote[\the \c@footnoteANote]{#1}}%
}

\makeatother

\author{%
  Hanfeng Chen%
  \thanksAAffil{Uppsala University and Center for Monetary Policy and Financial Stability (CeMoF).}%
  \and %
  Maria Elena Filippin%
  \footnotemarkAAffil[1]$^{\,,\,}$%
  \thanksANote{\textit{E-mail}: \texttt{mariaelena.filippin@nek.uu.se}}%
}

\date{June 12, 2025}

\begin{document}

\newgeometry{top=0.95in, bottom=0.95in, left=22mm, right=22mm}
\maketitle
\thispagestyle{empty}
\maketitle

\begin{abstract}
We analyze the risks to bank intermediation following the introduction of a central bank digital currency (CBDC) competing with commercial bank deposits as households’ source of liquidity. We revisit the result in the literature regarding the equivalence of payment systems introducing a collateral constraint on banks borrowing from the central bank. Comparing two equilibria with and without the CBDC, we find that even with this constraint, the central bank can ensure the same equilibrium allocation and price system by offering loans to banks. However, to access loans, banks must hold collateral at the expense of extending credit to firms. Thus, while the CBDC introduction has no real effects on the economy, it does not guarantee full neutrality as it affects banks’ business models. In a dynamic model extension, we examine the effects of an increase in the CBDC and show that the CBDC does not cause bank disintermediation or crowd out deposits but may foster an expansion of bank credit to firms.
\vspace{0.5cm}
\textbf{Keywords}: CBDC, collateral-constrained bank, equivalence, credit extension\blfootnote{This paper previously circulated under the title ``Unveiling the Interplay between Central Bank Digital Currency and Bank Deposits''. We are grateful to Mikael Carlsson, Daria Finocchiaro, Ulf Söderström, and Karl Walentin for their invaluable guidance and continuous support. For their useful feedback and comments, we also thank Itai Agur, Ragna Alstadheim, Lorenzo Burlon (discussant), Darrel Duffie, Rod Garratt, Ricardo Lagos, Cyril Monnet, Dirk Niepelt (discussant), Anna Rogantini Picco (discussant), and Joshua Weiss (discussant), as well as conference and seminar participants at Uppsala University, New York University, the 2nd PhD Workshop in Money and Finance at Sveriges Riksbank, the 2nd Conference on the Economics of CBDC by the Bank of Canada and Sveriges Riksbank, the CEPR-ECB Conference on the Macroeconomic Implications of CBDCs, the 16th Nordic Summer Symposium in Macroeconomics, and the Economics of Payments XIII Conference at the Oesterreichische Nationalbank. We are also grateful to the anonymous referee and to Editor Luc Laeven for their constructive comments, which helped improve the paper.}
\end{abstract}
\clearpage
\setcounter{page}{1}
\restoregeometry

\section{Introduction}

Digital currencies have been around for a while, but their potential significance in the global economy has increased recently due to a growing demand for digital payment methods for retail purposes and the gradual decline of the use of cash for transactions in many economies [see, e.g., \textcite{IMF2022}]. Besides the private digital means of payment currently in circulation, many central banks have been investigating the possibility of launching a central bank digital currency (CBDC). CBDCs are central bank liabilities denominated in an existing unit of account that serve as a medium of exchange and a store of value.\footnote{
As defined by the Committee on Payments and Market Infrastructures of the Bank for International Settlements [\textcite{CPMI2019}].
} If CBDCs were issued to retail customers (i.e., households), they would be a digital form of cash that share features with banknotes, as they are universally accessible but in a digital form.

Alongside an intense policy debate, a growing academic literature on the broader economic implications of CBDCs has emerged. A CBDC represents a novel payment alternative to cash and commercial bank deposits, with macroeconomic consequences that will affect both individuals and financial institutions. A primary concern for central banks when considering the issuance of a CBDC is the risk of the CBDC disintermediating the banking sector. If households substitute bank deposits \textit{en masse} for CBDC, it could potentially reduce the banks' ability to fund investments in the productive sector, thus resulting in overall negative economic effects. This paper analyzes the implications of the introduction of a retail CBDC, particularly its relationship with bank deposits.

The recent literature establishes an equivalence result between different payment systems. \textcite{Brunnermeier2019} consider a simplified scenario without reserves and resource cost of providing liquidity, while \textcite{Niepelt2022} includes a reserves layer and shows that introducing CBDC has no real effects on the economy if the private and the public sectors are equally efficient in operating payment systems. For this to happen, the central bank must refinance the banks at a lending interest rate that renders banks indifferent to the portfolio and policy changes.

\textcite{Niepelt2022} assumes that central bank loans are extended against no collateral. However, the collateral requirement imposed by central banks when lending to commercial banks is potentially important for how the introduction of a CBDC may affect the banking sector and the real economy.\footnote{
See, e.g., \textcite{Burlon2022} and \textcite{Williamson2022}.
} In practice, central banks lend to commercial banks (i.e., discount window lending) against collateral to support the liquidity and stability of the banking system. The liquidity provided by central banks helps financial institutions manage their liquidity risks efficiently. These loans are issued at an administered discount rate and must be collateralized to the satisfaction of the issuing central bank. In the euro area, banks can make use of the marginal lending facility, which enables banks to obtain overnight liquidity from the European Central Bank against sufficient eligible assets. In the United States, the Federal Reserve offers different types of discount window credit, which must be collateralized. The discount window mechanism has become increasingly important after the Global Financial Crisis. In this paper, we revisit this equivalence result in the literature and explore its implication in terms of financial disintermediation by introducing a financial friction for central bank lending to banks (i.e., the collateral requirement).\footnote{
In Appendix \ref{equiv2}, we revisit the equivalence result considering different degrees of substitutability between CBDC and bank deposits (i.e., imperfect substitutability).
}

This paper addresses the potential risk of bank disintermediation following the introduction of a CBDC. To address this concern, we build on \textcite{Niepelt2022} and develop a model with a CBDC and bank deposits, adding a collateral requirement for central bank lending to banks. The framework is an extension of the model by \textcite{Sidrauski1967} that embeds a banking sector, bank deposits, government bonds, reserves, and a CBDC into the baseline real business cycle model. Households value goods and the liquidity services that deposits and CBDC provide. Non-competitive banks invest in capital, reserves, and government bonds and fund themselves through either deposits or borrowing from the central bank. Firms produce using labor and physical capital.\footnote{
We assume households supply a fixed amount of labor.} 
Finally, the consolidated government collects taxes, invests in capital, lends to banks against collateral, and issues CBDC and reserves.

We study the introduction of a CBDC in two distinct ways. In the first part of the paper, we revisit the result in the literature regarding the equivalence of payment systems, comparing equilibria with and without CBDC. We derive the conditions under which there is an equivalence between the equilibria, even in the presence of a collateral constraint. Confirming the results in \textcite{Niepelt2022}, we find that the government can ensure equivalence as long as (i) CBDC and deposits are perfect substitutes, (ii) the resource cost per unit of effective real balances is the same for CBDC and deposits, and (iii) the central bank offers loans to banks at an interest rate that renders them indifferent to the introduction of the CBDC. However, due to the collateral constraint on central bank lending, the loan rate we derive is lower than the one in \textcite{Niepelt2022}. Intuitively, when banks are collateral-constrained, a lower interest rate is needed to incentivize them to take up the sufficient level of central bank loans that ensures equivalence. Furthermore, we show that while the central bank can compensate for the banks' decreased deposit funding and insulate their profits, banks nevertheless reduce loans to firms to meet collateral requirements. In other words, although the government can ensure the same equilibrium allocation and price system after introducing CBDC, it cannot guarantee ``full neutrality'' as the banks' business models change. Under certain conditions, the government’s capital holdings also expand, i.e., the government assumes a greater role in intermediating credit in the economy.

In the second part of the paper, we use our model to simulate the introduction of CBDC as a shock and study its dynamic implications. In this case, we are interested in the economy's responses when the government \emph{does not} seek to ``sterilize" the effect of CBDC, therefore, we do not impose any of the conditions under which the equivalence result holds. We simulate a gradual and near-permanent increase in CBDC to $5\%$ of steady-state output and find that CBDC in circulation expands the banks' balance sheets. This is because banks, which we model as having market power, react to the introduction of CBDC by reducing the (opportunity) cost of deposits (i.e., narrowing the interest spread on deposits), and thus experience deposit inflows. Together with banks' uptake of collateralized central bank loans, the introduction of CBDC increases banks' loans to firms. Hence, CBDC does not lead to bank disintermediation or crowd out deposits but expands the banks' credit extension in the economy.\\

\noindent
\textbf{Related literature}. \hspace{0.5mm} Our study contributes to the literature on the equivalence of payment systems. Existing works by \textcite{Brunnermeier2019} and \textcite{Niepelt2022} propose a compensation mechanism in which the households' shift from deposits to CBDC can be offset by central bank lending to banks. However, these models abstract from the collateral constraint for central bank lending that is common in practice. Notably, \textcite{Piazzesi2022} show that when banks are required to hold liquid assets to back their deposits and face asset management costs, the equivalence between alternative payment systems breaks down, even if banks can be refinanced directly by the central bank. In light of this, we revisit the equivalence result by incorporating a collateral constraint for banks. We derive a new central bank lending rate that ensures the equivalence of payment systems but depends on the restrictiveness of the collateral requirement. Our findings reveal that the more restrictive the collateral constraint, the lower the loan rate the central bank must post.

Moreover, our work contributes to the recent literature examining the impact of the introduction of CBDC on commercial banks. For instance, \textcite{Chiu2023} develop a micro-founded general equilibrium model based on the framework of \textcite{Lagos2005} calibrated to the U.S. economy and find that a CBDC expands bank intermediation when the price of CBDC falls within a certain range while leading to disintermediation if its interest rate exceeds the upper limit of that range. \textcite{Whited2023} build a dynamic banking model assuming that banks rely on deposits and wholesale funding and that the latter can potentially substitute deposit loss. Depending on whether the CBDC pays interest or not, the synergies between deposits and lending can attenuate the impact of the CBDC. In another work, \textcite{Keister2022} consider a competitive market and show that a deposit-like CBDC tends to crowd out bank deposits but, at the same time, increases the aggregate stock of liquid assets in the economy, promoting more efficient levels of production and exchange and ultimately raising welfare. Similarly, \textcite{Paul2024} develop a dynamic stochastic general equilibrium (DSGE) model with monopolistic banks and find that introducing a CBDC enhances household liquidity and limits banks' market power over deposits, but it can also diminish bank lending and profitability, highlighting a critical welfare trade-off in the CBDC design. Our dynamic model extension further illustrates that the introduction of CBDC need not disintermediate banks, given their pricing power in the deposit market and likely reactions to CBDC.

Lastly, our study belongs to the strand of literature examining the relationship between collateral requirements on banks and CBDC. In recent work, \textcite{Burlon2022} construct a quantitative euro area DSGE model, where banks must post government bonds as collateral to borrow from the central bank. They investigate the transmission channels of the issuance of CBDC to bank intermediation, finding a bank disintermediation effect with central bank financing replacing deposits, and government bonds displacing reserves and loans. Along similar lines, \textcite{Assenmacher2021} use a DSGE model to investigate the macroeconomic effects of CBDC when the central bank administers the CBDC rate and collateral, and quantity requirements. Their findings indicate that a more ample supply of CBDC reduces bank deposits, while stricter collateral or quantitative constraints reduce welfare but can potentially contain bank disintermediation. The latter effect is particularly true when the elasticity of substitution between bank deposits and CBDC is low. \textcite{Williamson2022}, on the other hand, explores the effects of the introduction of CBDC using a model of multiple means of payment. In his model, the CBDC is a more efficient payment instrument than cash, but it lengthens the central bank's balance sheet, creating collateral scarcity in the economy. Different from these works, our study investigates the implications of CBDC issuance on bank intermediation using a real business cycle model that is closely connected to the baseline macroeconomic workhorse model, building on \textcite{Niepelt2022} and embedding a collateral requirement for central bank lending to banks.\\

The rest of the paper is organized as follows. Section \ref{model} describes the model. Section \ref{equiv} revisits and discusses the analysis of the equivalence between payment systems. Section \ref{geneq} characterizes the general equilibrium in which the household holds CBDC and deposits and discusses the dynamic effects of an increase in the CBDC. Section \ref{conclusion} concludes.

\section{Model with CBDC and collateral-constrained banks} \label{model}
The model is based on \textcite{Niepelt2022} and describes an economy with a banking sector and CBDC in the absence of nominal rigidities. CBDC and deposits provide direct utility. We depart from that framework by considering a collateral constraint for banks when borrowing from the central bank. There is a continuum of mass one of homogeneous infinitely-lived households who own a succession of two-period-lived banks and of one-period-lived firms. The consolidated government determines the monetary and fiscal policy.

\subsection{Households}
The representative household wants to maximize the discounted felicity function $\mathcal{U}$, which is increasing, strictly concave, and satisfies the Inada conditions. Subject to its budget constraint, equation (\ref{hhbc}), the household takes wages, $w_t$; returns on asset $i$, $R^i_t$; profits, $\Pi_t$; and taxes, $\tau_t$ as given and solves
\begin{align}
    &\max_{{\{c_t, k_{t+1}^h, m_{t+1}, n_{t+1}\}}_{t=0}^\infty} \mathbb{E}_0 \sum_{t=0}^\infty \beta^t \mathcal{U}(c_t, z_{t+1}) \nonumber\\
    \text{s.t. \qquad} &c_t + k_{t+1}^h + m_{t+1} + n_{t+1} +\tau_t = w_t\bar{q} + \Pi_t +k_t^h R^k_t + m_t R^m_t + n_t R^n_t, \label{hhbc}\\
    &k_{t+1}^h, m_{t+1}, n_{t+1} \geq 0 \nonumber,
\end{align}
where $\beta \in (0,1)$ is the positive discount factor; $c_t$ denotes household consumption at date $t$; $k_{t+1}^h$ is capital at date $t+1$; and $z_{t+1} = z(m_{t+1}, n_{t+1})$ are effective real balances carried from date $t$ to $t+1$. Effective real balances are a function of both CBDC, $m_{t+1}$, and bank deposits, $n_{t+1}$.\footnote{
The household values liquidity, as suggested by the \textit{money in the utility function} specification. In this setting, it only matters that the household demands liquidity services, not why they do.
} We assume that the household inelastically supplies a constant amount of labor $\bar{q}$. The household consumes, pays taxes, invests in capital and real balances, out of wage income, distributed profits, and the gross return on the portfolio.

To express the first-order conditions in a more compact form, we define the risk-free rate as
\begin{align}
    R^f_{t+1} = \frac{1}{\mathbb{E}_t \Lambda_{t+1}}, \label{rf}
\end{align}
where $\Lambda_{t+1}$ is the household's stochastic discount factor:
\begin{align}
    \Lambda_{t+1} = \beta\frac{\mathcal{U}_c(c_{t+1}, z_{t+2})}{\mathcal{U}_c(c_t, z_{t+1})}. \label{Lambda}
\end{align}
Also, we define the liquidity premium, or interest spread, on asset $i$ as
\begin{align}
    \chi^i_{t+1} = 1 - \frac{R^i_{t+1}}{R^f_{t+1}}, \qquad i \in \{m, n\}.
    \label{spread}
\end{align}
The spread on asset $i$ denotes the household's opportunity cost of holding said asset. For example, a positive deposit spread shows the interest return that the household forgoes by holding deposits. The household is willing to accept a lower return on deposits due to the liquidity service they provide. We assume that the interest rates on CBDC and deposits are risk-free. Focusing on interior solutions for capital, CBDC, and deposits, the first-order conditions can be summarized as
\begin{align}
    k_{t+1}^h: \qquad  &1 = \mathbb{E}_t R^k_{t+1}\Lambda_{t+1}, \label{eek}\\
    m_{t+1}: \qquad &\mathcal{U}_c(c_t, z_{t+1})\chi^m_{t+1} = \mathcal{U}_z(c_t, z_{t+1}) z_m(m_{t+1}, n_{t+1}), \label{eem}\\
    n_{t+1}: \qquad &\mathcal{U}_c(c_t, z_{t+1})\chi^n_{t+1}= \mathcal{U}_z(c_t, z_{t+1}) z_n(m_{t+1}, n_{t+1}) \label{een}.
\end{align}

The household's first-order conditions have standard interpretations. The Euler equation for capital (\ref{eek}) dictates that the household saves in capital to the point where the marginal cost of saving in terms of consumption, $\mathcal{U}_c(c_t, z_{t+1})$, equals its expected discounted return, $\beta\mathbb{E}_t \mathcal{U}_c(c_{t+1}, z_{t+2}) R_{t+1}^k$. The first-order conditions for CBDC and deposits, equations (\ref{eem}) and (\ref{een}) respectively, show that the household demands the liquid asset $i \in \{ m, n \}$ to the point where its marginal benefit $\mathcal{U}_z(c_t, z_{t+1})z_{i}(m_{t+1},n_{t+1})/\mathcal{U}_c(c_t, z_{t+1})$ equals its opportunity cost in terms of foregone interest, $\chi^i_{t+1}$. Here, we only consider equilibria where CBDC and deposits circulate simultaneously. This is shown by equations (\ref{eem}) and (\ref{een}) both holding with equality. The co-existence of CBDC and deposits is ensured by the condition
\begin{align}
    z_m(m_{t+1}, n_{t+1}) \chi^n_{t+1} = z_n(m_{t+1}, n_{t+1}) \chi^m_{t+1}. \label{zmn}
\end{align}
Equation (\ref{zmn}) shows that, in equilibria where CBDC and deposits co-exist, the household allocates between them so that the marginal rate of substitution, $z_m(m_{t+1}, n_{t+1})/z_n(m_{t+1}, n_{t+1})$, equals the relative price, $\chi^m_{t+1}/\chi^n_{t+1}$. In equilibria where equation (\ref{zmn}) does not hold with equality, corner solutions arise in which only one payment instrument circulates. 

\subsection{Banks}
One of the often-cited reasons in the literature for introducing a CBDC is bank market power [see, e.g., \textcite{Andolfatto2021}, \textcite{Garratt2022}]. Specifically, banks offer lower deposit rates to extract rents, and households are willing to accept this as they value the liquidity service provided by deposits. A CBDC could compete with bank deposits, reducing banks' market power.

Our set-up is similar to \textcite{Niepelt2022} and assumes that each bank is a monopsonist in its regional deposit market, such that the household in a region can only access the regional bank. A bank lives for two periods, and at date $t$ issues deposits, $n_{t+1}$, and borrows from the central bank, $l_{t+1}$. It invests in reserves, $r_{t+1}$, government bonds, $b_{t+1}$, and capital, $k^b_{t+1}$.\footnote{Bank's capital is defined as $k^b_{t+1} = n_{t+1}+l_{t+1}-r_{t+1}-b_{t+1}$. Alternatively, the bank can invest in loans to firms that eventually fund physical capital accumulation.
} We interpret the productive capital held by banks as loans to non-financial firms. Therefore, in the following, we use the terms ``banks' capital investments'' and ``banks' loans to firms'' interchangeably. Without loss of generality, we abstract from bank equity.

We follow \textcite{Burlon2022} and assume that the bank is subject to a collateral requirement such that the loans they get from the central bank can not exceed a fraction $\theta_b$ of its government bond holdings. In this setting, government bonds are the only asset that can be pledged as collateral. For simplification, we abstract away from interbank loans with collateral. Holding government bonds gives liquidity benefits to the bank since they can use their holdings to obtain funding from the central bank. In other words, the bank is willing to forego a spread on the risk-free rate because of the collateral benefits of holding government bonds. This ``convenience yield" of government bonds reflects the additional benefits the bank derives from holding these bonds beyond their financial yield. Therefore, government bonds are remunerated at a slightly lower rate than the risk-free rate.

The operating costs in the retail payment system, $\nu$, represent the resource cost per unit of deposit funding that the bank incurs when transforming illiquid assets into liquid means of payment. Larger reserve holdings relative to deposits reduce the extent of liquidity transformation and therefore lower these costs. Following \textcite{Niepelt2024}, one microfoundation for these costs builds on a fire-sales narrative, following \textcite{Shleifer1992} and \textcite{Stein2012}. When, as a consequence of liquidity transformation, a bank lacks reserves to settle payments, it must sell capital (i.e., loans). Asymmetric information or market frictions depress the price of capital when other institutions find themselves in the same situation. Reserve holdings reduce the need for such costly capital liquidation, lowering the bank's operating costs.    
    
We interpret $\nu$ as a technological constraint rather than a regulatory one, and we treat it as exogenous from a policy perspective. We model $\nu$ as a decreasing function of the bank's reserve-to-deposit ratio, $\zeta_{t+1}$. We also allow $\nu$ to decrease with the stock of reserves and deposits of other banks, $\Bar{\zeta}_{t+1}$, so as to capture positive externalities of reserve holdings.\footnote{
\textcite{Niepelt2022} uses a cost function in the form $\nu + \omega_t(\zeta_{t+1},\Bar{\zeta}_{t+1})$, where $\nu$ is the resource cost per unit of deposit funding, and $\omega$ represents the bank's resource costs of liquidity substitution.} 
To simplify the analysis, we make some assumptions which imply that in equilibrium $\zeta_{t+1} = \Bar{\zeta}_{t+1}$, and reserves are strictly positive if and only if deposits are strictly positive: when a bank holds no deposits, its operating costs are null, and when all other banks have no deposits, the bank's operating costs are large but bounded. In this way, we rule out asymmetric equilibria in the bank's deposits and other banks' deposits. Otherwise, the operating cost function, $\nu(\zeta_{t+1},\Bar{\zeta}_{t+1})$, is strictly decreasing in both arguments, strictly convex, and satisfies $\nu_{\zeta \Bar{\zeta}}=0$ and $\nu_{\zeta \zeta} \geq \nu_{\Bar{\zeta} \Bar{\zeta}}$, as well as $ \lim_{\zeta_{t+1}\to 0} \nu_{\zeta} = \infty$.

The bank chooses the quantity of deposits and central bank loans, subject to the deposit funding schedule of the household.\footnote{
In the model, the central bank's loan funding schedule replicates the household's deposit funding schedule. This assumption plays a crucial role in the context of the equivalence analysis.
} Since the bank acts as a monopsonist in its regional deposit market, it takes the deposit funding schedule (rather than the deposit and the central bank loan rates) as given. The program of the bank at date $t$ reads
\begin{align}
    &\max_{n_{t+1}, l_{t+1}, r_{t+1}, b_{t+1}} \Pi^b_{1,t} + \mathbb{E}_t \left[ \Lambda_{t+1} \Pi^b_{2,t+1} \right] \nonumber \\
    \text{s.t. \qquad} &\Pi^b_{1,t} = -n_{t+1 } \nu(\zeta_{t+1},\Bar{\zeta}_{t+1}), \label{pi1}\\
    &\Pi^b_{2,t+1} = (n_{t+1}+l_{t+1}-r_{t+1}-b_{t+1})R^k_{t+1} \nonumber \\
    & \qquad \qquad + r_{t+1}R^r_{t+1} + b_{t+1}R^b_{t+1} - n_{t+1}R^n_{t+1} - l_{t+1}R^l_{t+1}, \label{pi2}\\
    &l_{t+1} \leq \theta_b \frac{b_{t+1}}{R^l_{t+1}}, \label{cc}\\
    & R^n_{t+1}, R^l_{t+1} \text{ perceived endogenous},\nonumber\\
    &n_{t+1}, l_{t+1}, b_{t+1} \geq 0 \nonumber,
\end{align}
where
\begin{align*}
    \zeta_{t+1} \equiv \frac{r_{t+1}}{n_{t+1}}, \qquad \Bar{\zeta}_{t+1} \equiv \frac{\Bar{r}_{t+1}}{\Bar{n}_{t+1}},
\end{align*}
and $\Pi^b_{1,t}$, $\Pi^b_{2,t+1}$ denote the cash flow generated in the first and second periods of the bank's operations, respectively.

We focus on interior solutions for deposits, loans, and government bonds, and we make use of the risk-free rate and the household's first-order condition for capital, equations (\ref{rf}) and (\ref{eek}), respectively. Also, we define the elasticity of the asset $i$ with respect to the rate of return on $i$ as
\begin{align}
    \eta_{i,t+1} = \frac{\partial i_{t+1}}{\partial R^i_{t+1}} \frac{R^i_{t+1}}{i_{t+1}}, \qquad i \in \{n, l\},
\end{align}
and the liquidity premia on central bank loans, reserves, and government bonds as in equation (\ref{spread}). Let $\gamma_t$ denote the Lagrange multiplier associated with the collateral constraint. 
The collateral constraint is binding in equilibrium, such that\footnote{
See Appendix \ref{sec:app:extracheck} for the conditions under which the collateral constraint binds. The intuition is that, with a non-binding collateral constraint, in equilibrium $\gamma_t = 0$ and, from the collateral constraint condition (\ref{cc}), $0 \leq \theta_b (b_{t+1}/R^l_{t+1})-l_{t+1}.$ However, from the government bonds optimality condition (not shown here), this violates the condition that $R^b_{t+1}<R^f_{t+1}$, so the collateral constraint must bind in equilibrium.
}
\begin{align*}
    \gamma_t>0, \qquad l_{t+1} =  \theta_b \frac{b_{t+1}}{R^l_{t+1}}.
\end{align*}
We can write the bank's optimality conditions as
\begin{align}
    n_{t+1}: \qquad &\chi^n_{t+1} -\left( \nu(\zeta_{t+1},\Bar{\zeta}_{t+1}) - \nu_{\zeta}(\zeta_{t+1},\Bar{\zeta}_{t+1}) \zeta_{t+1}\right)  = \frac{1}{\eta_{n,t+1}}\frac{R^n_{t+1}}{R^f_{t+1}} \label{optn},\\
    r_{t+1}: \qquad &-\nu_{\zeta}(\zeta_{t+1},\Bar{\zeta}_{t+1}) = \chi^r_{t+1} \label{optr},\\
    l_{t+1}: \qquad & \chi^l_{t+1} - \gamma_t\left(1+\frac{1}{\eta_{l,t+1}}\right) = \frac{1}{\eta_{l,t+1}}\frac{R^l_{t+1}}{R^f_{t+1}} \label{optl},\\
    b_{t+1}: \qquad & \gamma_t \frac{\theta_b}{R^l_{t+1}} = \chi^b_{t+1} \label{optb}, 
\end{align}
where the spreads on bonds and central bank loans, $\chi^b_{t+1}$ and $\chi^l_{t+1}$, are defined in the same way as the spreads on deposits and CBDC given by expression (\ref{spread}).

We first comment on the liability side of the bank's balance sheet, starting with deposits. The left-hand side of the equation (\ref{optn}) represents the marginal profit from issuing deposits, which is given by the difference between the bank's gain from the positive deposit liquidity premium and the marginal cost associated with increased deposit issuance. The right-hand side equals the marginal cost of inframarginal deposits, as higher deposit issuance is associated with an increased interest rate on deposits. Similarly, the condition for central bank loans, equation (\ref{optl}), states that the sum of the bank's marginal benefits of taking on more central bank loans and the gain coming from the positive loan liquidity premium should be equal to the marginal cost associated with central bank loans. In fact, higher loan holdings are associated with an increase in the interest rate on the central bank loans.

Turning now to the asset side of the bank's balance sheet, equation (\ref{optr}) equalizes the marginal benefit of reserves in the form of reduced operating costs with the bank's opportunity cost of reserves. Looking at equation (\ref{optb}), the optimal choice of government bonds is when the bank's marginal costs of bond holdings are equal to the loss coming from the bank's lower return with a positive spread on government bonds.

Combining equations (\ref{optn}) and (\ref{optr}) yields
\begin{equation}
\tag{\ref{optn}a}
    \chi^n_{t+1} - \big[ \nu(\zeta_{t+1},\Bar{\zeta}_{t+1})  +\chi^r_{t+1}\zeta_{t+1} \big] = \frac{1}{\eta_{n,t+1}}\frac{R^n_{t+1}}{R^f_{t+1}}.
\end{equation}
This implies that the bank's net benefit of issuing more deposits must equal the inframarginal cost of deposits. Combining equations (\ref{optl}) and (\ref{optb}) results in the relation
\begin{equation}
\tag{\ref{optl}a}
    \chi^l_{t+1} - \chi^b_{t+1}\frac{R^l_{t+1}}{\theta_b} \left(1+\frac{1}{\eta_{l,t+1}}\right) = \frac{1}{\eta_{l,t+1}}\frac{R^l_{t+1}}{R^f_{t+1}}.
\end{equation}
The marginal cost of taking on more central bank loans must equal the bank's net benefit of taking on more loans. This is given by the difference between the liquidity benefits given by the central bank loans and the marginal cost associated with the collateral constraint.

\subsection{Firms}
Neoclassical firms live for one period and rent capital, $k_t$, and labor, $\ell_t$, to produce the output good to maximize the profit, $\Pi^f_t$. The representative firm takes wages, $w_t$; the rental rate of capital, $R^k_t + \delta -1$; and the good price as given and solves
\begin{equation*}
\begin{split}
    &\max_{k_t, \ell_t} \Pi^f_t\\
    \text{s.t. \qquad} &\Pi^f_t = f(k_t,\ell_t) - k_t(R^k_t + \delta -1) - w_t\ell_t,
\end{split}
\end{equation*}
where $f$ is the neoclassical production function and $\delta$ is the rate of capital depreciation. The first-order conditions read
\begin{align}
    k_t: \qquad &f_k(k_t,\ell_t) = R^k_t +\delta -1, \label{kappa}\\
    \ell_t: \qquad &f_l(k_t,\ell_t) = w_t, \label{ell}
\end{align}
where $f_k$ and $f_l$ are the first-derivatives of $f$ with respect to capital and labor, respectively. 

\subsection{Consolidated government}
The consolidated government collects taxes, lends to the bank against collateral, invests in capital, $k^g_{t+1}$, and issues CBDC and reserves. The government budget constraint reads
\begin{align}
    k^g_{t+1} +l_{t+1} -b_{t+1} - m_{t+1} -r_{t+1} =& k^g_t R^k_t +l_t R^l_t-b_t R_t^b -m_t R^m_t -r_t R^r_t \nonumber\\
    & +\tau_t -m_{t+1} \mu -r_{t+1} \rho \label{gbc},
\end{align}
where $\mu$ and $\rho$ are the unit resource costs of issuing (and managing) CBDC and reserves payments, respectively.

\subsection{Market clearing}
Market clearing in the labor market requires that the firm's labor demand equals the household's (inelastic) labor supply:
\begin{align}
    \ell_t = \bar{q} \label{mclab}.
\end{align}
Market clearing for capital requires that the firm's demand for capital equals capital holdings of the household, the bank, and the government:
\begin{align}
    k_t = k_{t}^h + (n_t+l_t-r_t-b_t) + k_t^g \label{mck}. 
\end{align}
Profits distributed to the household must equal the sum of the bank and firm profits:
\begin{align}
    \Pi_t = \Pi^b_{1,t} + \Pi^b_{2,t} + \Pi^f_t \label{totp}.
\end{align}
By Walras' law, market clearing on labor and capital markets and the budget constraints of the household, bank, firm, and consolidated government imply market clearing on the goods market.

To derive the aggregate resource constraint for the economy, we plug equation (\ref{totp}) into the household's budget constraint, equation (\ref{hhbc}), and we impose market-clearing conditions (\ref{mclab}) and (\ref{mck}). Then, in combination with the government's budget constraint, equation (\ref{gbc}), the resulting expression is the aggregate resource constraint:
\begin{align}
    k_{t+1} = f(k_t,\bar{q})+k_t(1-\delta)-c_t-\left(m_{t+1}\mu+n_{t+1}\left(\nu(\zeta_{t+1},\Bar{\zeta}_{t+1}) + \zeta_{t+1}\rho\right)\right) \label{aggrc}. 
\end{align}

\section{Revisiting the equivalence of payment systems}\label{equiv}
In this section, we compare equilibria with and without CBDC, revisiting the result in the literature regarding the equivalence of payment systems. 

Existing works by \textcite{Brunnermeier2019} and \textcite{Niepelt2022} suggest a compensation mechanism where the households' shift from deposits to CBDC can be offset by central bank lending to banks. We build on these insights by incorporating a collateral constraint for central bank lending, as specified in equation (\ref{cc}) from the bank's problem. We assume perfect substitutability between CBDC and deposits, such that effective real balances, $z_{t+1}$, are a weighted sum of the two payment instruments:\footnote{
The assumption that an interest-bearing CBDC and deposits are close substitutes is common in the literature [see, e.g., \textcite{Andolfatto2021} and \textcite{Whited2023}]. See Appendix \ref{equiv2} for the equivalence analysis in the case of CBDC and deposits as imperfect substitutes.
}
\begin{align}
    z_{t+1} = \lambda m_{t+1}+n_{t+1},\label{funz}
\end{align}
where $\lambda \geq 0$ represents the benefits of CBDC relative to deposits. Besides reflecting the liquidity benefits or ease of use of CBDC, $\lambda$ can also capture factors such as privacy considerations and payment security.

The following proposition is formally proved in Appendix \ref{sec:app:eqprof}.\\

\noindent
\textbf{Proposition 1}. \hspace{0.5mm} \textit{Consider a policy that implements an equilibrium with deposits and reserves. The central bank introduces a new payment instrument, CBDC, which is a perfect substitute for deposits. There exists another policy and equilibrium with less deposits and reserves, a positive amount of CBDC, central bank loans, and government bonds which act as collateral for central bank loans, a different ownership structure of capital, additional taxes on the household, and otherwise the same equilibrium allocation and price system. In this equilibrium, the public and private sectors are equally efficient in providing liquidity to the household:
\begin{align}
    \frac{\mu}{\lambda} = \nu(\zeta_{t+1}, {\zeta}_{t+1}) + \zeta_{t+1}\rho, \label{cond}
\end{align}
and the central bank lends to the bank at an interest rate equal to
\begin{align}
    R^l_{t+1} = \frac{R^n_{t+1} +\nu(\zeta_{t+1},\zeta_{t+1}) R^f_{t+1} -\zeta_{t+1}R^r_{t+1}}{(1-\zeta_{t+1})\left(1+\frac{R^k_{t+1}-R^b_{t+1}}{\theta_b} \right)} \label{eqrate2}.
\end{align}%
$\blacksquare$}

Condition (\ref{cond}) stipulates that the resource cost the government pays to provide one unit of real balances through CBDC equals the corresponding cost for deposits, which is the sum of the operating cost incurred by the bank and the resource cost associated with reserves incurred by the government.\footnote{
This condition is in line with the likely implementation scheme in which central banks issue CBDCs to the public using the existing commercial banks' deposit distributing systems [see, e.g., \textcite{BIS2023} for the results of the 2022 BIS survey on central bank digital currencies and cryptocurrencies].} 
In other words, condition (\ref{cond}) requires that public and private provision of liquidity are equally cost-efficient. The government budget constraint is unaffected by the new policy as long as condition (\ref{cond}) holds, and the central bank lends to the bank at the loan rate given by equation (\ref{eqrate2}). This loan rate also ensures that the market value of taxes on the household is zero, implying that the household budget constraint is unaffected by the new policy.\footnote{
The additional taxes on the household are set to compensate for the changes in bank profits collected by the household.} 
Consequently, the household’s optimal sequences of consumption and real balances remain unchanged. Lastly, the loan rate also ensures that the market value of changes in the bank's profits is zero. It follows that, by offering loans and setting the appropriate interest rate, the central bank can insulate the bank’s profits, such that the composition of the bank's portfolio continues to be optimal.

The loan rate we derive is lower than the one in \textcite{Niepelt2022} due to the extra term $[1+(R^k_{t+1}-R^b_{t+1})/\theta_b] >0$ in equation (\ref{eqrate2}).\footnote{
From the household's problem, we know that $R^k_{t+1} \leq R^f_{t+1}$, assuming that the rate of return on capital is not risky, we can approximate $R^k_{t+1} \simeq R^f_{t+1}$. We also know that for the bank there is a collateral premium associated with holding government bonds, thus $R^b_{t+1} < R^f_{t+1}$. It follows that the extra term is positive.
} It follows that when the bank is subject to a collateral constraint for central bank lending, the central bank must offer a lower interest rate than in a no-collateral constraint scenario. The intuition is that when the bank is not collateral-constrained, it can borrow as much as it wants from the central bank. With a collateral requirement, the central bank needs to offer a lower lending rate to incentivize the bank to borrow the same quantity as in the absence of the constraint, such that it remains indifferent to the introduction of the CBDC.\footnote{We abstract from any social cost associated with central bank lending to banks.} 
Note that in limit scenarios, the central bank loan rate we derive coincides with the one in \textcite{Niepelt2022}, e.g., if the returns on capital and bonds are identical, or if $\theta_b$ approaches infinity.

Importantly, in our setting, the central bank loan rate depends on how restrictive the collateral constraint is: the tighter the constraint is (the lower $\theta_b$), the lower the lending rate the central bank needs to offer (the lower $R^l_{t+1}$). Intuitively, when the bank is more collateral-constrained, a lower interest rate is needed to incentivize the bank to take up a sufficient level of central bank loans to ensure equivalence.

\begin{figure}[!htbp]
\centering
\caption{Bank's balance sheets}
\vskip 0.8em
\begin{minipage}{\textwidth}
    \centering
    \makebox[\textwidth]{\includegraphics[width=0.8\textwidth]{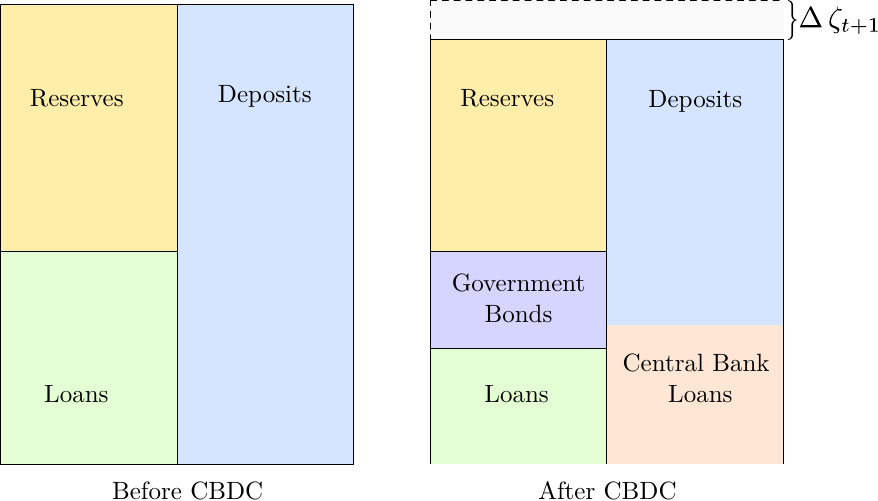}}
\end{minipage}
\vskip 1em
\begin{minipage}{\textwidth}
    \footnotesize
    \textit{Note}: This figure shows the bank's balance sheet composition before and after the CBDC introduction. Following the new policy, the balance sheet shrinks by $\Delta \,\zeta_{t+1}$, where $\Delta >0$ is the decrease in deposit holdings, and $\zeta_{t+1}$ represents the reserve-to-deposit ratio.
\end{minipage}
\label{fig:bs}
\end{figure}

Some final remarks are worth noting. First, after the new policy, to access central bank loans, the bank must hold government bonds as collateral, thus crowding out loans to firms, as shown in Figure \ref{fig:bs} reporting the bank's balance sheet breakdown before and after the introduction of the CBDC. This is not the case in the original results of \textcite{Niepelt2022}, where the reduction in the bank's balance sheet does not affect the amount of loans the bank extends to firms, as the fall in the bank liabilities is exactly offset by fewer reserves. Note that, after the new policy, while the household balance sheet remains unchanged as they only change the composition of the asset holdings, the bank's balance sheet shrinks by $\Delta \,\zeta_{t+1}$, where $\Delta >0$ is the decrease in deposit holdings, and $\zeta_{t+1}$ represents the reserve-to-deposit ratio. Second, the fall in bank loans to firms equals the uptake of bonds, and is decreasing in the collateral constraint $\theta_b$ (under reasonable calibrations). In other words, given a less restrictive collateral constraint (higher $\theta_b$), the crowding out of bank loans is less severe (see Appendix \ref{sec:app:eqprof} for the details).

Given the above considerations, the following corollary follows from Proposition 1:\\

\noindent
\textbf{Corollary 1}. \hspace{0.5mm} \textit{In the presence of the collateral constraint, the central bank loan rate in equation (\ref{eqrate2}) insulates bank profits but cannot prevent the crowding-out of the bank loans to firms. The reduction in bank loans equals the uptake of government bonds and is decreasing in the collateral constraint. Although the central bank can ensure equivalence in allocation and prices, it cannot guarantee ``full neutrality'' as the bank's business model is altered as a result of the portfolio and policy changes.}\\
\noindent
$\blacksquare$

\subsection{Discussion}
The equivalence result in the literature is akin to a Modigliani--Miller benchmark, suggesting that the introduction of CBDC will not affect the real economy through reduced credit provision, as long as the central bank offsets any disintermediation of banks. The idea is that when households shift from deposits to CBDC, the central bank can lend to banks to replace the lost deposits, maintaining the overall credit supply in the economy.

We extend this framework by introducing a collateral constraint on central bank lending, adding a realistic financial friction. The scope of this analysis is to contribute to the theoretical rationale for why the equivalence result may not fully hold. To access central bank loans, banks must hold government bonds as collateral. As a result, as households partially substitute deposits with CBDC, banks replace their lost funds with central bank loans, but to do so, they must redirect part of their reserves and loans to firms towards purchasing government bonds. This reshaping of bank portfolios alters their business model, even though the central bank can insulate profits and preserve aggregate capital or real production.

In particular, the reduction in bank loans to firms is compensated by either the government or the household. For instance, when CBDC and deposits provide equal utility to the household (i.e., $\lambda = 1$), the government alone covers the reduction in bank loans, effectively assuming part of the credit extension role. In practice, central banks are cautious about placing themselves in a position where they need to re-intermediate lost bank funding. For this reason, many central banks are designing CBDCs to limit disintermediation, for instance, by considering caps on CBDC wallets [see, e.g., \textcite{ecb2024c}].

In conclusion, in this Section, we revisit the equivalence result in the literature by introducing a collateral constraint on central bank lending and show that, while the central bank can replicate the original equilibrium allocation and insulate bank profits, the introduction of CBDC alters the size and composition of the bank balance sheet. While our focus is on characterizing the conditions and mechanisms behind the equivalence results under a collateral framework, we do not explore policies that improve upon the pre-CBDC equilibrium. Such welfare-based comparisons lie beyond the scope of the equivalence analysis, but could offer an interesting direction for future research.

\section{Dynamic effects of introducing CBDC}\label{geneq}
In Section \ref{equiv}, we establish the conditions under which equilibria with and without CBDC are equivalent, even in the presence of a collateral constraint for central bank lending to banks. In this section, we use our model to simulate the introduction of a CBDC as a shock and study its dynamic implications, more specifically, the potential threat to bank intermediation should the CBDC crowd out deposits. We are interested in the economy's responses in a case where the government \emph{does not} seek to ``sterilize'' the effects of introducing CBDC. In other words, we do not impose any of the conditions under which the equivalence results from the previous section hold. Specifically, the central bank loan rate does not necessarily follow the equivalent rate as in equation (\ref{eqrate2}), and the public and private provision of liquidity are not necessarily equally cost-efficient, i.e., condition (\ref{cond}) does not necessarily hold. 

\subsection{Functional forms and equilibrium conditions}
The functional form for real balances is represented by equation (\ref{funz}). We assume that the household has utility function of the form 
\begin{align}
    \mathcal{U}(c_t, z_{t+1})=\frac{\left((1-\iota)c_t^{1-\psi}+\iota z_{t+1}^{1-\psi}\right)^{\frac{1-\sigma}{1-\psi}}}{1-\sigma}, \label{hhu}
\end{align}
where $\iota>0$ is the utility weight of liquidity; $\sigma>0$ is the inverse intertemporal elasticity of substitution between bundles of consumption and real balances across times; and $\psi>0$ is the inverse intratemporal elasticity of substitution between consumption and real balances. The bank's operating cost function has the following form:
\begin{align}
    \nu(\zeta_{t+1},\bar{\zeta}_{t+1}) = \phi_{1}\zeta_{t+1}^{1-\varphi}+\phi_{2}\bar{\zeta}_{t+1}^{1-\varphi}, \label{bcost}
\end{align}
where $\phi_1,\phi_2\geq 0$ are the relative weights assigned to the bank's reserves-to-deposit ratio and to the other bank's ratio; and $\varphi>1$. Lastly, the firm has the standard Cobb-Douglas production function:
\begin{align}
     f(k_t,\ell_t) = k_t^{\alpha}\ell_t^{1-\alpha}, \label{CD}
\end{align}
where $\alpha$ is the capital share of output.

Given the functional form assumptions, we characterize the general equilibrium. First, knowing the household's utility functional form, we can rewrite the stochastic discount factor, equation (\ref{Lambda}) as
\begin{align*}
    \Lambda_{t+1} = \beta \frac{c_{t+1}^{-\sigma}\Omega_{t+1}}{c_t^{-\sigma}\Omega_t}, 
\end{align*}
so that the risk-free rate, equation (\ref{rf}), is given by
\begin{align*}
    R_{t+1}^f = \frac{1}{\mathbb{E}_t\left[\Lambda_{t+1}\right]}.
\end{align*}
The household's capital Euler equation (\ref{eek}) and the aggregate resource constraint (\ref{aggrc}) become, respectively,  
\begin{align}
    &1 = \mathbb{E}_t\left[\Lambda_{t+1}R_{t+1}^k\right], \label{eqk}\\
    &k_{t+1} = k_t^{\alpha}(\bar{q})^{1-\alpha}+k_t(1-\delta)-c_t-\left(m_{t+1}\mu+n_{t+1}\left(\nu(\zeta_{t+1},\zeta_{t+1}) + \zeta_{t+1}\rho\right)\right) , \label{eqaggrc}
\end{align}
where 
\begin{align}
    \Omega_t &= (1-\iota)^{\frac{1-\sigma}{1-\psi}}\left(1+\left(\frac{\iota}{1-\iota}\right)^{\frac{1}{\psi}}\left({\chi^n_{t+1}}\right)^{1-\frac{1}{\psi}}\right)^{\frac{\psi-\sigma}{1-\psi}}. \label{omega}
\end{align}
These equilibrium conditions closely parallel those of a standard real business cycle model. Unlike the standard model, however, the auxiliary variable $\Omega_t$ summarizes the impact of the household's preference for liquidity on consumption/savings choices. Moreover, the terms $m_{t+1}\mu$ and $n_{t+1}\left(\nu(\zeta_{t+1},\zeta_{t+1}) + \zeta_{t+1}\rho\right)$ in the resource constraint show the societal costs of providing liquidity to the household incurred by the government and banks, respectively. The societal costs of providing CBDC, $m_{t+1}\mu$, are simply the product of the amount of CBDC in circulation and the government's unit cost of supply (and managing) CBDC. On the other hand, the societal costs of bank deposits, $n_{t+1}\left(\nu(\zeta_{t+1},\zeta_{t+1}) + \zeta_{t+1}\rho\right)$, include both the bank's direct resource cost, $n_{t+1}\nu(\zeta_{t+1},\zeta_{t+1})$ and the government's cost of issuing reserves, $n_{t+1}\zeta_{t+1}\rho$, which are used to back up deposit issuance. 

We combine the household's first-order condition for deposits, equation (\ref{een}), with the expression for real balances, equation (\ref{funz}), to derive the deposit demand:
\begin{align}
    n_{t+1} = c_t \left(\frac{\iota}{1-\iota} \frac{1}{\chi^n_{t+1}}\right)^{\frac{1}{\psi}} -\lambda m_{t+1}.  \label{eqn}
\end{align}
We combine the bank first-order conditions for deposits (\ref{optn}) and reserves (\ref{optr}) to derive the expressions for the equilibrium deposit spread, $\chi_{t+1}^n$:
\begin{align}
    \chi_{t+1}^n = \frac{(\phi_{1}\varphi+\phi_{2})\zeta_{t+1}^{1-\varphi}}{1-\psi\frac{n_{t+1}}{z_{t+1}}}, \label{eqspreadn} 
\end{align}
and the bank's optimal reserves-to-deposits ratio, $\zeta_{t+1}$, which depends on the spread on reserves, $\chi_{t+1}^r$, 
\begin{align}
    \zeta_{t+1} = \left(\frac{\chi_{t+1}^r}{\phi_1(\varphi-1)}\right)^{-\frac{1}{\varphi}}. \label{zeta}
\end{align}
Given the real balances functional form assumption, equation (\ref{funz}), from the household's problem equation (\ref{zmn}) we derive the CBDC spread as
\begin{align}
    \chi_{t+1}^m = \lambda \chi_{t+1}^n. \label{eqzmn}
\end{align}
Note that the spread on reserves is derived from equation (\ref{spread}).

From the binding collateral constraint, equation (\ref{cc}), we derive the bank's demand for government bonds as
\begin{align}
    b_{t+1} = \frac{l_{t+1}R_{t+1}^l}{\theta_b}. 
\end{align}
Combining the bank's optimality conditions for central bank loans and government bonds, equations (\ref{optl}) and (\ref{optb}), we derive the bank's demand for central bank loans. We restrict our attention to the case where this demand is non-negative, i.e.,\footnote{
To derive equation (\ref{eqloan}), we assume that the central bank's loan supply schedule replicates the household's deposit demand schedule, i.e., 
$\partial l_{t+1}/\partial R^l_{t+1} = \partial n_{t+1}/\partial R^n_{t+1}$.
} 
\begin{align}
      l_{t+1} = \label{eqloan} 
    \begin{cases}
      \left(\chi^l_{t+1} -\chi^b_{t+1} \frac{R^l_{t+1}}{\theta_b}\right) \left(\frac{\theta_b}{\chi^b_{t+1} R^f_{t+1}+ \theta_b}\right) \frac{z_{t+1}}{\psi \chi^n_{t+1}} & \text{if $\chi^l_{t+1} -\chi^b_{t+1} \frac{R^l_{t+1}}{\theta_b}\geq 0$},\\
      0 & \text{if $\chi^l_{t+1} -\chi^b_{t+1} \frac{R^l_{t+1}}{\theta_b}< 0$}.
    \end{cases}  
\end{align}

Finally, from the firm's optimality conditions, equations (\ref{kappa}) and (\ref{ell}), we derive the return on capital and the real wage:
\begin{align}
    R_{t}^k &= 1-\delta +\alpha\left(\frac{k_{t}}{\bar{q}}\right)^{\alpha-1}, \label{rk}\\
    w_t&=(1-\alpha)\left(\frac{k_t}{\bar{q}}\right)^{\alpha}. \label{w}
\end{align}

\subsection{Shock}
After characterizing the general equilibrium, we aim to investigate the potential threat to bank intermediation should the CBDC crowd out deposits. We address this concern by studying the economy's responses to an increase in the supply of CBDC. We follow \textcite{Burlon2022} and assume that the central bank issues CBDC according to a policy rule which stipulates that CBDC is equal to a fraction of steady-state output, $y$:
\begin{align}
    m_{t+1} = \theta^m_t y, \label{cbdcrule}
\end{align}
where the share of CBDC, $\theta^m_t$ is time-varying.

In practice, the initial introduction of CBDC is likely to be gradual and permanent. To analyze the effects of such a change, we follow \textcite{chen} and model a near-permanent, gradual increase in the CBDC share using an AR(2) process
\begin{align}
    \Delta \theta_t^m = \rho_1^{\theta}\Delta \theta_{t-1}^m-\rho_2^{\theta}(\theta_{t-1}^m-\theta^m)+e_t, \label{ar2}
\end{align}
where $\rho_1^{\theta}$ and $\rho_2^{\theta}$ are persistence parameters; $\theta^m$ is the steady-state share of CBDC; and $e_t$ is the exogenous shock.

\subsection{Calibration}
The model is quarterly, and we calibrate it to the U.S. economy. We use variables without subscripts to denote their steady-state values. Table \ref{cal_params} summarizes the baseline calibration.

\subsubsection{Households}
The household's discount factor, $\beta$, is set to the standard value of $0.99$. We set the fixed labor supply, $\bar{q}$, to $1/3$. We assume that the household perceives CBDC and deposits as equally useful, i.e., $\lambda=1$. We set the inverse intertemporal elasticity of substitution, $\sigma$, to $0.5$. We assume that consumption and liquidity services are complements. Therefore, the inverse intratemporal elasticity of substitution between the two, $\psi$, is set higher than $\sigma$ and equal to $0.55$, aligned with the calibration in \textcite{Niepelt2022}. We calibrate the utility weight of liquidity, $\iota$ to $0.016$ to match the ratio of liquidity to output of $1.04$, in line with \textcite{kaplan} and \textcite{bayer}. 

\subsubsection{Banks and firms}
Following \textcite{Niepelt2022}, we set the bank operating cost parameter $\varphi$ to $2.3048$. We assume the internal and external costs of lack of reserves are identical, i.e., $\phi_2=\phi_1=\phi$. We then set $\phi$ to $0.0003$ to achieve a reserves-to-deposits ratio of $0.25$, as in \textcite{Niepelt2022}. The production sector is standard. The capital share of output, $\alpha$, and the rate of capital depreciation, $\delta$, are set to $1/3$ and $0.025$, respectively.

\subsubsection{Government}
We follow \textcite{Niepelt2022} and set the government's marginal cost of providing reserves, $\rho$, to $0.0004$. We assume the government is equally efficient in providing CBDC as reserves, i.e., we set the marginal cost of CBDC to $\mu=0.0004$. For simplicity, we assume that central bank reserves, central bank loans, and government bonds are non-interest-bearing (in real terms). We calibrate the persistence parameters in equation (\ref{ar2}), representing the supply of CBDC, to ensure a gradual increase in the CBDC share that reaches $5\%$ after approximately 20 quarters, i.e., $\rho_1^{\theta}=0.845$ and $\rho_2^{\theta}=0$. The longer-term CBDC share of $5\%$ is consistent with \textcite{Abad2023}, who analyze the transitional dynamics in scenarios with steady-state take-up of CBDC equal to between $4\%$ and $7\%$ of GDP. Lastly, we set the haircut on government bonds to $0.5\%$, which implies a collateral haircut $\theta_b=0.995$.

\begin{table}[!htbp]
\centering
    \begin{threeparttable}
    \caption{Model parameters}
    \label{cal_params}
    \begin{tabular}{l*{4}{c}}
    \toprule \toprule
    \multicolumn{1}{c}{Parameter} & \multicolumn{1}{c}{Value} & \multicolumn{1}{c}{Source/Motivation} & \multicolumn{1}{c}{Description} \\
    \midrule 
    \multicolumn{4}{l}{Households} \\
    $\beta$    & $0.99$  & Standard   & Discount factor \\ 
    $\bar{q}$    & $1/3$  & Standard   & Fixed labor supply \\ 
    $\lambda$  & $1$     & Assumption & Relative CBDC benefit \\
    $\sigma$   & $0.5$   & Assumption & Risk aversion \\
    $\psi$     & $0.55$   & $\psi>\sigma$ & Inv. elasticity of sub. $c_t$ and $z_{t+1}$ \\ 
    $\iota$    & $0.016$ & $z/y=1.04$ & Liquidity utility weight \\ 
    \midrule 
    \multicolumn{4}{l}{Banks} \\
    $\varphi$  & $2.3048$ & \textcite{Niepelt2022} & Operating cost \\ 
    $\phi_1$   & $0.0003$ & $\zeta=0.25$& Operating cost  \\ 
    $\phi_2$   & $0.0003$ & $\phi_1=\phi_2$ & Operating cost\\ 
    \midrule 
    \multicolumn{4}{l}{Firms} \\
    $\alpha$   & $1/3$   & Standard & Capital share of output \\
    $\delta$   & $0.025$ & Standard & Capital depreciation rate \\
    \midrule 
    \multicolumn{4}{l}{Government} \\
    $\rho$     & $0.0004$ & \textcite{Niepelt2022} & Reserves cost \\
    $\mu$      & $0.0004$  & $\mu=\rho$ & CBDC cost \\    
    $R^l$      & $1.0$    & Assumption & Central bank loans returns \\
    $R^r$      & $1.0$    & Assumption & Reserves return \\
    $R^b$      & $1.0$    & Assumption & Government bonds return \\
    $\rho_1^{\theta}$ & $0.845$ & Assumption & CBDC supply persistence \\
    $\rho_2^{\theta}$ & $0$ & Assumption & CBDC supply persistence \\
    $\theta_b$ & 0.995    & Haircut on bonds $0.5\%$ & Collateral haircut \\
    \bottomrule
    \end{tabular}
    \end{threeparttable}
    \vskip 0.5em
    \begin{minipage}{\textwidth}
        \footnotesize
        \textit{Note}: This table reports the calibrated model parameters.
    \end{minipage}
\end{table}

\subsection{Impulse responses}
In this section, we keep the interest rates on reserves, central bank loans, and government bonds constant at their steady-state levels. The impulse responses are reported as percentage or basis point deviations from steady-state. Note that, in the baseline, CBDC deviations are reported in absolute values because in the steady-state there is no CBDC. 

\begin{figure}[!htbp]
\caption{Impulse responses to a near-permanent increase in CBDC share from zero to 5\%}
\begin{minipage}{\textwidth}
    \centering
    \makebox[\textwidth]{\includegraphics[width=1\textwidth]{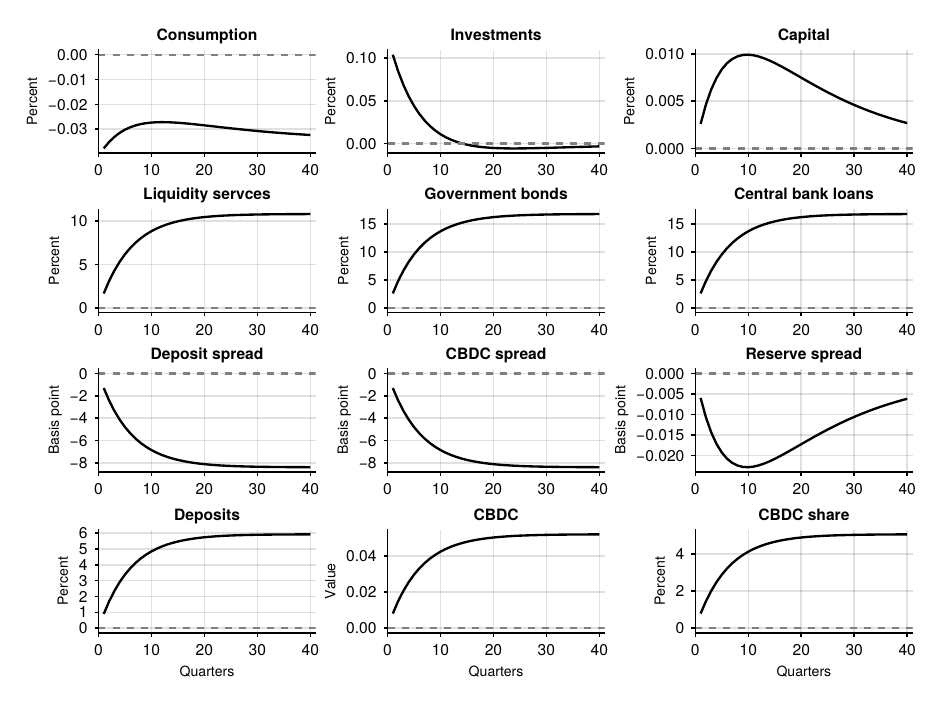}}
\end{minipage}
\vskip 1em
\begin{minipage}{\textwidth}
    \footnotesize
    \textit{Note}: This figure shows the impulse responses to a gradual and near-permanent increase in CBDC from zero to $5\%$ of steady-state output. The initial increase in the CBDC share is about 0.8 percentage points.
\end{minipage}
\label{fig:fig_perm_em}
\end{figure}

Figure \ref{fig:fig_perm_em} illustrates the impulse responses to a gradual and near-permanent increase in the CBDC share, $\theta^m_t$, that reaches $5\%$ of steady-state output after about 20 quarters.\footnote{
The initial increase in CBDC as a share of steady-state output is about $0.8$ percentage point.
} This is a simple way of simulating the transition from a steady-state without CBDC to a new state with a constant, positive amount of CBDC. Recall that this new state differs from the equivalent equilibrium with CBDC we study in Section \ref{equiv}. In this analysis, we do not impose any of the conditions that ensure that equilibria with and without CBDC are equivalent. We are instead interested in a situation where the government supplies CBDC but does not seek to ``sterilize'' its effects. 

On impact, the increase in the CBDC share decreases the equilibrium deposit spread, pinned down by equation (\ref{eqspreadn}). The reason is twofold. First, the reserve spread falls, which increases the bank's liquidity ratio and reduces the cost of issuing deposits, given by the term $(\phi_{1}\varphi+\phi_{2})\zeta_{t+1}^{1-\varphi}$. This puts downward pressure on the deposit spread. Second, the fraction of deposits in total liquidity services falls, which reduces the markup that the bank can charge over the marginal cost of issuing deposits, represented by the term $1/[1-\psi(n_{t+1}/ z_{t+1})]$. In other words, the introduction of CBDC reduces the market power of the bank.

Since deposits and CBDC are perfect substitutes, the government ensures that both assets circulate by setting the return on CBDC such that equation (\ref{eqzmn}) holds. The government ensures that the opportunity costs of holding deposits and CBDC are equalized, adjusting for the relative benefit of CBDC. As a result, the CBDC spread decreases by the same magnitude as the deposit spread.

Equation (\ref{eqn}) shows that the household's demand for deposits decreases in both the deposit spread and the amount of CBDC in circulation. Importantly, deposit demand increases because the effect of the declining deposit spread is stronger than the crowding-out effect of increased CBDC. The increase in deposits is accompanied by an increase in the bank's demand for central bank loans, enabling the bank to expand its balance sheet. According to equation (\ref{eqloan}), this expansion is primarily driven by the greater availability of total liquid assets and the reduced deposit spread. To secure central bank loans, the bank must post collateral, leading to an increase in government bond holdings.

More aggregate liquidity also increases the marginal utility of consumption. Recalling that liquidity services and consumption are complements (i.e., $\psi>\sigma$), the increase in liquidity raises the marginal utility of any level of consumption. This can be seen more directly from the variable $\Omega_t$, given by equation (\ref{omega}), which summarizes the impact of the household’s preference for liquidity on the marginal utility of consumption, given by $c_t^{-\sigma}\Omega_t$. $\Omega_t$ depends solely, and negatively, on the deposit spread. Its impulse response (not plotted here) is therefore the inverse of the response of the deposit spread. A persistently falling deposit spread then raises the marginal utility of any level of consumption and, in turn, leads to a moderately lower level of consumption over time. 

\begin{figure}[!htbp]
\caption{Impulse responses to a near-permanent increase in CBDC share from zero to 5\%}
\begin{minipage}{\textwidth}
    \centering
    \makebox[\textwidth]{\includegraphics[width=1\textwidth]{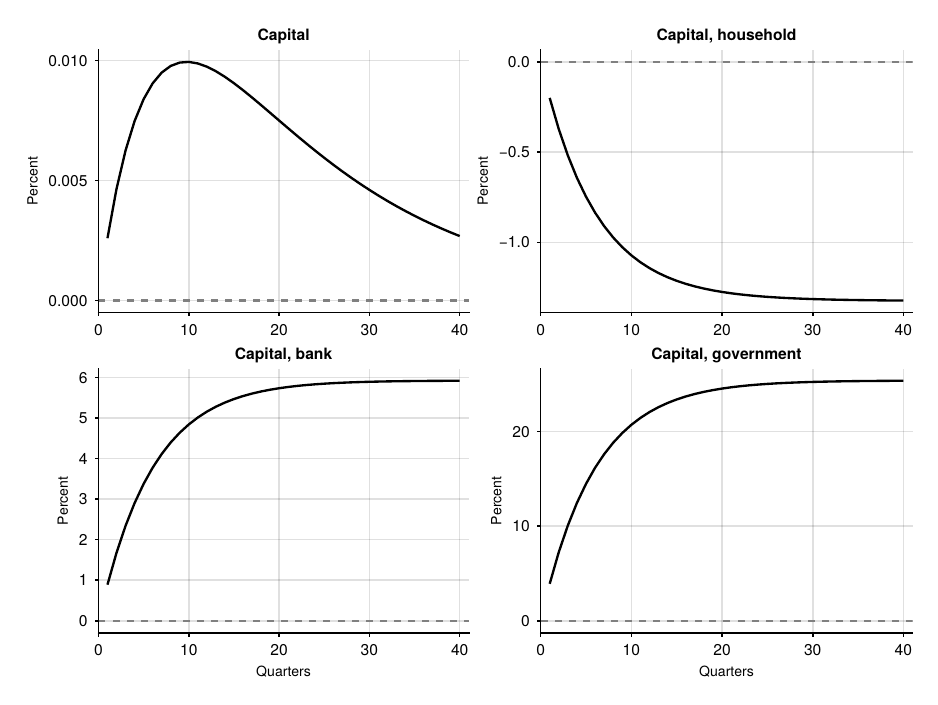}}
\end{minipage}
\vskip 1em
\begin{minipage}{\textwidth}
    \footnotesize
    \textit{Note}: This figure shows the impulse responses to a gradual and near-permanent increase in CBDC from zero to $5\%$ of steady-state output. The initial increase in the CBDC share is about 0.8 percentage points.
\end{minipage}
\label{fig:k_decomp}
\end{figure}

The aggregate capital in the economy is the sum of capital held by households, banks, and the government. Increased household savings in liquid assets come at the cost of decreased capital holdings. Figure \ref{fig:k_decomp} shows the aggregate capital breakdown by components. Household capital gradually declines to about $1.9\%$ lower than before the introduction of CBDC. On the other hand, both the bank and government capital holdings increase, making up for the fall in household capital and resulting in a minor increase in aggregate capital and the associated increase in investments. The bank's larger balance sheet leads to increased capital investments. As shown in Figure \ref{fig:k_decomp}, the bank raises its capital holdings persistently up to around $6\%$ over time. The intuition behind the bank's capital increase is that the bank expands credit intermediation to firms. In other words, introducing CBDC does not cause bank disintermediation but expands bank activity.\footnote{
See Appendix \ref{sec:app:dynamics} for the alternative specification in which the CBDC share is described by an AR(1) process.}

Lastly, some remarks regarding welfare are worth noting. A natural question is whether the introduction of CBDC is welfare-improving. While explicit welfare calculations are beyond the scope of this paper, some intuition can be provided. We have thus far assumed that the banking sector is non-competitive. Consequently, the resulting equilibrium allocation without CBDC is sub-optimal, as banks exercise market power by setting the deposit spread above their marginal cost of providing deposits. As a result, the quantity of deposits, and thus aggregate liquidity, is lower than the level that would support an optimal allocation. Thus, the introduction of CBDC, with the resulting increase of aggregate liquidity, is likely to be welfare-improving along the transition.

\subsection{Robustness checks}
Uncertainty regarding the household's perception of CBDC usefulness would be important for any practical implementation of CBDC. Therefore, we first test the robustness of our results by changing the relative benefit of CBDC. Second, given the central role of the pledgeability of bonds for our equivalence results, we test for different magnitudes for the haircuts on bonds.  

First, we change the relative benefit of CBDC, $\lambda$. Figures \ref{fig:irf_lam05} and \ref{fig:irf_lam15} in the Appendix \ref{sec:app:robustness} show the impulse responses to a near-permanent increase in CBDC share, $\theta_t^m$, from zero to $5\%$, when $\lambda$ is $0.5$ and $1.5$, respectively. Comparing these responses to the main specification in Figure \ref{fig:fig_perm_em}, we see that the results remain qualitatively the same. The magnitudes of the responses are smaller when the relative benefit of CBDC is smaller, i.e., when $\lambda=0.5$. This is because each unit of CBDC contributes less to total liquidity as CBDC comes into circulation, which in turn means the share of deposits in total liquidity, $n_{t+1}/z_{t+1}$, is at a higher level immediately after the shock than in the baseline. Intuitively, this means that for the same increase in CBDC, its competitive impact on the bank is smaller. Then, according to equation (\ref{eqspreadn}), the bank does not have to lower its deposit spread as much as in the baseline. The opposite is true when $\lambda$ is higher.

Next, Figures \ref{fig:irf_thb999} and \ref{fig:irf_thb985} in the Appendix \ref{sec:app:robustness} show the impulse responses to a temporary increase in CBDC share when bond haircuts are at $0.1\%$ ($\theta_b=0.999$) and $1.5\%$ ($\theta_b=0.985$), respectively. We see that the changes in bond pledgeability do not alter the responses in shape or magnitude.

\section{Conclusion} \label{conclusion}
This paper investigates the potential risk to bank intermediation when introducing a CBDC that competes with commercial bank deposits as the household's source of liquidity. We model CBDC and banks subject to a collateral requirement and analyze the static and dynamic effects of the CBDC.

We revisit the result in the literature regarding the equivalence of payment systems, comparing equilibria with and without CBDC. We derive the conditions under which there is an equivalence between the equilibria, even in the presence of a collateral constraint. Confirming the results in \textcite{Niepelt2022}, we find that the government can ensure equivalence as long as (i) CBDC and deposits are perfect substitutes, (ii) the resource cost per unit of effective real balances is the same for CBDC and deposits and (iii) the central bank offers loans to banks at an interest rate that renders them indifferent to the introduction of the CBDC. However, because of the collateral constraint on central bank lending, the loan rate we derive is lower than the one in \textcite{Niepelt2022}. Intuitively, when the banks are collateral-constrained, a lower interest rate is needed to incentivize them to take up the sufficient level of central bank loans that ensures equivalence. Furthermore, we show that while the central bank can compensate for the banks' decreased deposit funding and insulate their profits, banks nevertheless reduce loans to firms in order to meet collateral requirements. Put differently, although the government can ensure the same equilibrium allocation and price system after introducing CBDC, it cannot guarantee ``full neutrality'' as the banks' business models change. Under certain conditions, the government’s holdings of capital also expand, i.e., the government assumes a greater role in intermediating credit in the economy.

We also use our model to simulate the introduction of CBDC as a shock and study its dynamic implications. Here, we are interested in the responses of the economy in a case where the government \emph{does not} seek to “sterilize” the effect of CBDC, therefore, we do not impose any of the conditions under which the equivalence results from the previous section hold. We simulate a gradual and near-permanent increase in CBDC to $5\%$ of steady-state output and find that CBDC in circulation expands the banks' balance sheets. This is because banks, which we model as having market power, react to the introduction of CBDC by reducing the interest spread on deposits, and thus experience deposit inflows. Together with the banks' uptake of collateralized central bank loans, the introduction of CBDC increases banks' investments in capital. Hence, a CBDC need not lead to bank disintermediation or crowd out deposits, but could instead expand the banks' credit extension in the economy.

\newpage

\nocite{*}
\printbibliography[
    heading=bibintoc,
    title=References,
    ]

\newpage
\appendix

\section{Condition under which the collateral constraint binds}
\label{sec:app:extracheck}
Assuming interior solutions, the bank's optimality conditions for loans and bonds are, respectively:
\begin{align}
    \mathbb{E}_t \left[ \Lambda_{t+1}(R^k_{t+1} - R^l_{t+1} - l_{t+1}\frac{\partial R^l_{t+1}}{\partial l_{t+1}}) \right] &= \gamma_t \left(1 + \theta_b \frac{b_{t+1}}{{R^l}^{2}_{t+1}} \frac{\partial R^l_{t+1}}{\partial l_{t+1}} \right), \tag{A1}\\
    \mathbb{E}_t \left[ \Lambda_{t+1} (R^k_{t+1} -R^b_{t+1}) \right] &= \gamma_t \frac{\theta_b}{R^l_{t+1}}, \tag{A2}
\end{align}
where $\gamma_t$ denotes the Lagrange multiplier associated with the collateral constraint. Subtracting the condition for bonds from the one for loans:
\begin{align}
\label{eq2}
    \mathbb{E}_t \left[ \Lambda_{t+1}(R^b_{t+1} - R^l_{t+1}  - l_{t+1}\frac{\partial R^l_{t+1}}{\partial l_{t+1}}) \right] = \gamma_t \left( 1- \frac{\theta_b}{R^l_{t+1}} +\theta_b \frac{b_{t+1}}{{R^l}^{2}_{t+1}} \frac{\partial R^l_{t+1}}{\partial l_{t+1}} \right). \tag{A3}
\end{align}
To define the sign of the RHS, recall that $\theta_b \in [0,1]$, and since the rate of return on reserves is positive, and we assume interior solutions, all the terms are positive.

We define the elasticity of the central bank's loans with respect to their rate of return as
\begin{align}
     \eta_{l,t+1}=\frac{\partial l_{t+1}}{\partial R^l_{t+1}} \frac{R^l_{t+1}}{l_{t+1}}, \tag{A4}
\end{align}
such that we can rewrite the last term on the LHS as
\begin{align}
    \frac{1}{\eta_{l,t+1}} R^l_{t+1}. \tag{A5}
\end{align}
Expression (\ref{eq2}) says that the collateral constraint is binding if
\begin{align}
    R^b_{t+1} - R^l_{t+1} > \frac{1}{\eta_{l,t+1}} R^l_{t+1}. \tag{A6}
\end{align}
Rearranging gives:
\begin{align}
    R^b_{t+1} > R^l_{t+1} +\frac{1}{\eta_{l,t+1}}R^l_{t+1}. \tag{A7}
\end{align}
We can conclude that the collateral constraint is binding if the sum of the cost of borrowing from the central bank and the bank's cost of taking on more loans is cheaper than the return the bank gets from holding government bonds:\footnote{We replicated the same analysis in the setting by \textcite{Burlon2022}, and we got an analogous result.}
\begin{align*}
    \gamma_t>0, \qquad l_{t+1} =  \theta_b \frac{b_{t+1}}{R^l_{t+1}} \qquad \text{ iff } R^b_{t+1} > R^l_{t+1} +\frac{1}{\eta_{l,t+1}}R^l_{t+1}.
\end{align*}

\section{Equivalence} \label{sec:app:eqprof}
For convenience, we repeat Proposition 1 as in Section \ref{equiv} and prove it formally.\\

\noindent
\textbf{Proposition 1}. \hspace{0.5mm} \textit{Consider a policy that implements an equilibrium with deposits and reserves. The central bank introduces a new payment instrument, CBDC, which is a perfect substitute for deposits. There exists another policy and equilibrium with less deposits and reserves, a positive amount of CBDC, central bank loans, and government bonds which act as collateral for central bank loans, a different ownership structure of capital, additional taxes on the household, and otherwise the same equilibrium allocation and price system.\\
$\blacksquare$}
\\

Indicate the new equilibrium by circumflexes, and suppose that deposit holdings decrease by a magnitude of $\Delta$ from the initial equilibrium, i.e. $\hat{n}_{t+1} - n_{t+1} = -\Delta$. Suppose also that real balances, the aggregate capital stock, and the reserves-to-deposits ratio remain unchanged in the new equilibrium, i.e., 
\begin{align*}
    \hat{z}_{t+1} = z_{t+1}, \quad \hat{k}_{t+1} = k_{t+1}, \quad \hat{\zeta}_{t+1} = \zeta_{t+1}.
\end{align*}
The above implies the following changes in the other equilibrium quantities:\footnote{
To guarantee the non-negativity of deposits, capital holdings, and reserves, $\Delta$ must not be too large. Specifically, we impose
$$\Delta \leq n_{t+1}, \qquad \zeta_{t+1}\Delta \leq r_{t+1}, \qquad \left(1-\frac{1}{\lambda}\right)\Delta \leq k^g_{t+1}, \qquad \left(1-\frac{1}{\lambda}\right)\Delta \geq -k_{t+1}^h.$$
}
\begin{align*}
    \hat{m}_{t+1} - m_{t+1} = \frac{1}{\lambda}\Delta, &\qquad \hat{r}_{t+1} - r_{t+1} = -\zeta_{t+1}\Delta, \\
    \hat{l}_{t+1} = (1-\zeta_{t+1})\Delta, &\qquad \hat{b}_{t+1} = \frac{\hat{l}_{t+1}R^l_{t+1}}{\theta_b}, \\
    \hat{k}_{t+1}^h - k_{t+1}^h = \left(1-\frac{1}{\lambda}\right)\Delta, &\qquad \hat{k}^g_{t+1} - k^g_{t+1} = -\left(1-\frac{1}{\lambda}\right)\Delta +\hat{b}_{t+1}. 
\end{align*}

First, we show that the new policy has no real effects on the economy, given an appropriate level of interest rate on central bank loans. Note that, before the implementation of the new policy, the cash flows generated in the first and second periods of the bank's operations are given by equations (\ref{pi1}) and (\ref{pi2}), respectively. Recalling that in equilibrium $\zeta_{t+1} = \Bar{\zeta}_{t+1}$, the changes in bank profits at dates $t$ and $t + 1$ are, respectively:
\begin{align}
    \hat{\Pi}^b_{1,t} - \Pi^b_{1,t} &= \Delta\left(\nu(\zeta_{t+1}, \zeta_{t+1}) \right), \label{d_pi1} \tag{B1}\\
    \hat{\Pi}^b_{2,t+1} - \Pi^b_{2,t+1} &= \Delta \left( R^n_{t+1} -\zeta_{t+1}R^r_{t+1} -(1-\zeta_{t+1})\left(1+\frac{R^k_{t+1}-R^b_{t+1}}{\theta_b} \right)R^l_{t+1}\right). \label{d_pi2} \tag{B2}
\end{align}
Let $\hat{T}_{1,t}$ be a tax on the household at date $t$ that compensates for the reduced bank losses:
\begin{align}
    \hat{T}_{1,t} &= \hat{\Pi}^b_{1,t} - \Pi^b_{1,t} = \Delta\left(\nu(\zeta_{t+1}, \zeta_{t+1})  \right). \label{T1} \tag{B3}
\end{align}
We denote $\hat{T}_{2,t+1}$ as a tax at date $t+1$ that compensates for the change in the household's portfolio return as well as for the change in bank profits that the household collects at date $t+1$:\footnote{
Given the household's trade-offs between CBDC and deposits, expression (\ref{zmn}), it follows that: $\lambda/R_{t+1}^m = R_{t+1}^n-[1-(1/\lambda)]R_{t+1}^f$.
}
\begin{align}
     \hat{T}_{2,t+1} =\Delta\left[\left(1-\frac{1}{\lambda}\right)R_{t+1}^k+\frac{R_{t+1}^m}{\lambda}-\zeta_{t+1}R^r_{t+1} -(1-\zeta_{t+1})\left(1+\frac{R^k_{t+1}-R^b_{t+1}}{\theta_b} \right)R^l_{t+1}\right]. \label{AT2} \tag{B4}
\end{align}
Let $\mathcal{T}_t=\hat{T}_{1,t} + \mathbb{E}_t \Lambda_{t+1} \hat{T}_{2,t+1}$ denote the market value of taxes at date $t$. Substituting the two expressions for taxes, equations (\ref{T1}) and (\ref{AT2}), and using conditions from the household’s optimization problem, we can rewrite $\mathcal{T}_t$ as
\begin{align*}
    \mathcal{T}_t = \Delta \left[\nu(\zeta_{t+1}, \zeta_{t+1})  + \frac{R^n_{t+1} -\zeta_{t+1}R^r_{t+1} -(1-\zeta_{t+1})\left(1+\frac{R^k_{t+1}-R^b_{t+1}}{\theta_b} \right)R^l_{t+1}}{R^f_{t+1}}\right].  
\end{align*}
In order for the new policy to have no real effects on the economy, the market value of taxes must be zero. This is true if the central bank posts a loan rate equal to equation (\ref{eqrate2}):
\begin{align*}
    R^l_{t+1} = \frac{R^n_{t+1} +\nu(\zeta_{t+1},\zeta_{t+1})R^f_{t+1} -\zeta_{t+1}R^r_{t+1}}{(1-\zeta_{t+1})\left(1+\frac{R^k_{t+1}-R^b_{t+1}}{\theta_b} \right)}.    
\end{align*}
We denote the market value of the changes in bank profits at date $t$ as $\mathcal{P}_t = (\hat{\Pi}^b_{1,t} - \Pi^b_{1,t}) + \mathbb{E}_t \Lambda_{t+1}( \hat{\Pi}^b_{2,t+1} - \Pi^b_{2,t+1}).$
Plugging in the expressions for changes in bank profits, equations (\ref{d_pi1}) and (\ref{d_pi2}), and using the definition of the risk-free rate from the household's problem, $\mathcal{P}_t$ reads
\begin{align*}
    \mathcal{P}_t\ &= \Delta\left(\nu(\zeta_{t+1}, \zeta_{t+1}) \right) \\
    & \hspace{0.5cm} + \frac{1}{R^f_{t+1}} \Delta \left( R^n_{t+1} -\zeta_{t+1}R^r_{t+1} -(1-\zeta_{t+1})\left(1+\frac{R^k_{t+1}-R^b_{t+1}}{\theta_b} \right)R^l_{t+1}\right),
\end{align*}
which is equal to zero given equation (\ref{eqrate2}). It follows that if the central bank offers an interest rate on central bank loans according to equation (\ref{eqrate2}), the market values of the taxes and of the changes in bank profits are zero.

Next, we show that the government's dynamic and intertemporal budget constraints continue to be satisfied with the new policy. Before the implementation of the new policy, the government budget constraint at time $t$ reads:
\begin{align}
    k^g_{t+1} -m_{t+1} -r_{t+1} = k^g_t R^k_t -m_t R^m_t -r_t R^r_t+\tau_t -m_{t+1} \mu -r_{t+1} \rho. \label{G_t} \tag{B5}
\end{align}
The government budget constraint at time $t$ with the new policy and changes is
\begin{align*}
    \hat{k}^g_{t+1} +\hat{l}_{t+1} -\hat{m}_{t+1}  -\hat{r}_{t+1} -\hat{b}_{t+1} &= k^g_t R^k_t -m_t R^m_t -r_t R^r_t+\tau_t \\
    & \hspace{0.5cm}-\hat{m}_{t+1} \mu -\hat{r}_{t+1} \rho +\hat{T}_{1,t}.
\end{align*}
Rearranging, simplifying, and collecting terms:
\begin{align}
    k^g_{t+1} -m_{t+1} -r_{t+1} +\Delta \left(\frac{\mu^m}{\lambda} -\left(\nu(\zeta_{t+1}, \zeta_{t+1}) + \rho \zeta_{t+1} \right) \right) &= k^g_t R^k_t -m_t R^m_t -r_t R^r_t+\tau_t \nonumber \\
    & \quad -m_{t+1} \mu -r_{t+1} \rho. \label{G_t_hat} \tag{B6}
\end{align}
The government budget constraints before and after the intervention at time $t$, equations (\ref{G_t}) and (\ref{G_t_hat}), are identical as long as 
\begin{align*}
    \frac{\mu}{\lambda} = \nu(\zeta_{t+1}, {\zeta}_{t+1}) + \zeta_{t+1}\rho.
\end{align*}
That is, when the public and private sectors are equally efficient in providing liquidity to the household, the government’s budget constraint at date $t$ is unaffected by the changes in allocation.

\noindent
Similarly, before the new policy, the government budget constraint at time $t+1$ reads 
\begin{align}
    k^g_{t+2} -m_{t+2} -r_{t+2} &= k^g_{t+1} R^k_{t+1} -m_{t+1} R^m_{t+1} -r_{t+1} R^r_{t+1} \nonumber\\
    & \quad +\tau_{t+1} -n_{t+2}\theta_{t+1} -m_{t+2} \mu -r_{t+2} \rho. \label{G_t+1} \tag{B7}
\end{align}
With the new policy and changes, the government budget constraint at time $t+1$ becomes
\begin{align*}
    k^g_{t+2} -m_{t+2} -r_{t+2} &= \hat{k}^g_{t+1} R^k_{t+1} + \hat{l}_{t+1} R^l_{t+1} -\hat{m}_{t+1} R^m_{t+1} -\hat{r}_{t+1} R^r_{t+1} -\hat{b}_{t+1} R^b_{t+1}\\
    & \quad +\tau_{t+1} -n_{t+2}\theta_{t+1} -m_{t+2} \mu -r_{t+2} \rho +\hat{T}_{2,t+1}.
\end{align*}
Rearranging, simplifying, and collecting terms:
\begin{align}
    k^g_{t+2} -m_{t+2} -r_{t+2} = &\hat{k}^g_{t+1} R^k_{t+1} + \hat{l}_{t+1} R^l_{t+1} -\hat{m}_{t+1} R^m_{t+1} -\hat{r}_{t+1} R^r_{t+1} -\hat{b}_{t+1} R^b_{t+1} \nonumber \\
    & +\tau_{t+1} -n_{t+2}\theta_{t+1} -\hat{r}_{t+1} R^r_{t+1} - m_{t+2} \mu -r_{t+2} \rho +\hat{T}_{2,t+1}. \label{G_t+1_hat} \tag{B8}
\end{align}
Using the expression for the central bank loan rate we derived, equation (\ref{eqrate2}), it follows that the government budget constraints before and after the intervention at time $t+1$, equations (\ref{G_t+1}) and (\ref{G_t+1_hat}), are the same. In other words, the central bank loan rate ensuring that the market values of taxes and changes in bank profits are zero, also ensures that the government budget constraint at time $t+1$ is unaffected by the changes in allocation. 

We claimed initially that the proposed intervention does not change the price system. In this case, the firm's optimal production decisions and profits are unchanged. Lastly, we must show that the modified bank's portfolio is still optimal. Before the intervention, the bank's choice set is determined by the cost function, the household's stochastic discount factor, rates on returns on capital and reserves, and the deposit funding schedule. The new policy leaves unchanged the cost function, the stochastic discount factor, and the rates on returns on capital and reserves. After the intervention, as the household holds more CBDC, there is a modified deposit funding schedule, together with a central bank loan funding schedule. The central bank needs to post an appropriate loan funding schedule to induce the non-competitive bank to go along with the equivalent balance sheet positions as before the intervention. Subject to this schedule, the bank chooses loans that make up for the reduction in funding from the household, net of reserves, at the same effective price. The central bank chooses a loan funding schedule which mirrors the deposit funding schedule and posts the loan rate as in equation (\ref{eqrate2}).

\subsection{Equivalence with imperfect substitutability between payment systems}\label{equiv2}
In Section \ref{equiv} we consider CBDC and deposits as perfectly substitutable for the household. However, some works in the literature consider imperfect substitutability between the two instruments [see, e.g., \textcite{Agur2022}, \textcite{Bacchetta2022}, \textcite{Barrdear2022}, \textcite{Burlon2022}, and \textcite{Kumhof2021}]. We now assume a constant elasticity of substitution (CES) functional form for the household's real balances: 
\begin{align}
    z_{t+1}(m_{t+1},n_{t+1}) = \left( \lambda m^{1-\epsilon}_{t+1} + n^{1-\epsilon}_{t+1} \right)^\frac{1}{1-\epsilon},\label{funz2} \tag{B9}
\end{align}
where $\lambda \geq 0$ represents the benefits of CBDC relative to deposits, and $\epsilon \geq 0$ is the inverse elasticity of substitution between payment instruments.\\

\noindent
\textbf{Proposition 2}. \hspace{0.5mm} \textit{Consider a policy that implements an equilibrium with deposits and reserves. The central bank introduces a new payment instrument, CBDC, which is an imperfect substitute for deposits. There does not exist another policy and equilibrium that guarantees the same equilibrium allocation and price system.}\\
\noindent
$\blacksquare$

Suppose again that deposit holdings decrease by a magnitude of $\Delta$ from the initial equilibrium and that real balances, the aggregate capital stock and the reserves-to-deposits ratio remain unchanged in the new equilibrium. This implies the same changes in equilibrium quantities as we have seen in Appendix \ref{sec:app:eqprof}, except for the CBDC. Due to the imperfect substitutability between CBDC and deposits, in order for real balances to remain unchanged, the quantity of CBDC must change according to 
\begin{align*}
    \hat{m}_{t+1} - m_{t+1} = \left[\frac{1}{\lambda}\left(n_{t+1}^{1-\epsilon}-\hat{n}_{t+1}^{1-\epsilon}\right)+m_{t+1}^{1-\epsilon}\right]^{\frac{1}{1-\epsilon}} - m_{t+1}.
\end{align*}
We define taxes at dates $t$ and $t+1$, equations (\ref{T1}) and (\ref{AT2}) respectively, as in Appendix \ref{sec:app:eqprof}, as well as the market values of taxes, $\mathcal{T}_t = \hat{T}_{1,t} + \mathbb{E}_t \Lambda_{t+1} \hat{T}_{2,t+1}$. The central bank loan rate ensuring that the market value of taxes is zero is:\footnote{
When deriving the central bank loan rate, we use expression (\ref{zmn}) for the household's trade-off between CBDC and deposits. Given the CES functional form assumption (\ref{funz2}) it follows that: $m_{t+1}/n_{t+1} = \left(\lambda\chi_{t+1}^n/\chi_{t+1}^m\right)^\frac{1}{\epsilon}$.
} 
\begin{align}
    R_{t+1}^l = \frac{\mathcal{A}_t R_{t+1}^n-\zeta_{t+1}R_{t+1}^r+\left(\nu(\zeta_{t+1},\zeta_{t+1}) +1-\mathcal{A}_t\right)R_{t+1}^f}{(1-\zeta_{t+1})\left(1+\frac{R^k_{t+1}-R^b_{t+1}}{\theta_b} \right)}, \label{eqrate3} \tag{B10}
\end{align}
where 
\begin{align*}
    \mathcal{A}_t = \lambda\left(\frac{\hat{m}_{t+1} - m_{t+1}}{\Delta}\right)\left(\frac{n_{t+1}}{m_{t+1}}\right)^{\epsilon}. 
\end{align*}
Consider the changes in bank profits at dates $t$ and $t+1$ as given from equations (\ref{d_pi1}) and (\ref{d_pi2}), respectively. We can check whether the market value of the changes in bank profits, $\mathcal{P}_t = (\hat{\Pi}^b_{1,t} - \Pi^b_{1,t}) + \mathbb{E}_t \Lambda_{t+1}( \hat{\Pi}^b_{2,t+1} - \Pi^b_{2,t+1})$, also reduces to zero given the central bank loan rate in expression (\ref{eqrate3}). It turns out this is not true. In particular, after making the appropriate substitutions, the market value of changes in bank profits reads:
\begin{align*}
    \mathcal{P}_t = \mathbb{E}_t \frac{1}{R^f_{t+1}} \Delta\left(R_{t+1}^n-\mathcal{A}_t R_{t+1}^n-\left(1-\mathcal{A}_t\right)R_{t+1}^f\right). 
\end{align*}
Note that if there were perfect substitutability between CBDC and deposits (i.e., $\epsilon=0$), as in the case studies in Section \ref{equiv}, $\mathcal{A}_t$ equals 1, and the market value of the changes in bank profits reduces to zero. It follows that, in case of imperfect substitutability between CBDC and deposits, the central bank lending rate that renders the market value of taxes zero does not result in changes in bank profits being zero. In other words, the central bank cannot make the bank indifferent to the competition from CBDC. In fact, a change in the bank's profitability implies that the new policy does not guarantee the same equilibrium allocation as before, implying that the introduction of CBDC has real effects on the economy. In \textcite{Brunnermeier2019}, one condition for equivalence to hold is that CBDC and deposits are minimally substitutable, such that their marginal liquidity contribution is unchanged. When the two instruments are imperfect substitutes, as in the case under study, the marginal rate of substitution is not constant, and equivalence is not guaranteed.

\section{Dynamics} \label{sec:app:dynamics}
\renewcommand{\thetable}{C.\arabic{table}}
\renewcommand{\thefigure}{C.\arabic{figure}}
\setcounter{table}{0}
\setcounter{figure}{0}

We compare the baseline specification with the gradual and near-permanent increase in CBDC to one in which the CBDC share is described  by a more standard AR(1) process, which simulates a temporary increase: 
\begin{align}
    \theta^m_t =  (1-\rho^{\theta})\theta^m + \rho^{\theta} \theta^m_{t-1} + e_t,  \label{ar1} \tag{C1}
\end{align}
where the persistence of the process $\rho^{\theta}$ is set to $0.9$.

\begin{figure}[!htbp]
\caption{Impulse responses to a temporary increase in CBDC share from zero to 5\%}
\begin{minipage}{\textwidth}
    \centering
    \makebox[\textwidth]{\includegraphics[width=1\textwidth]{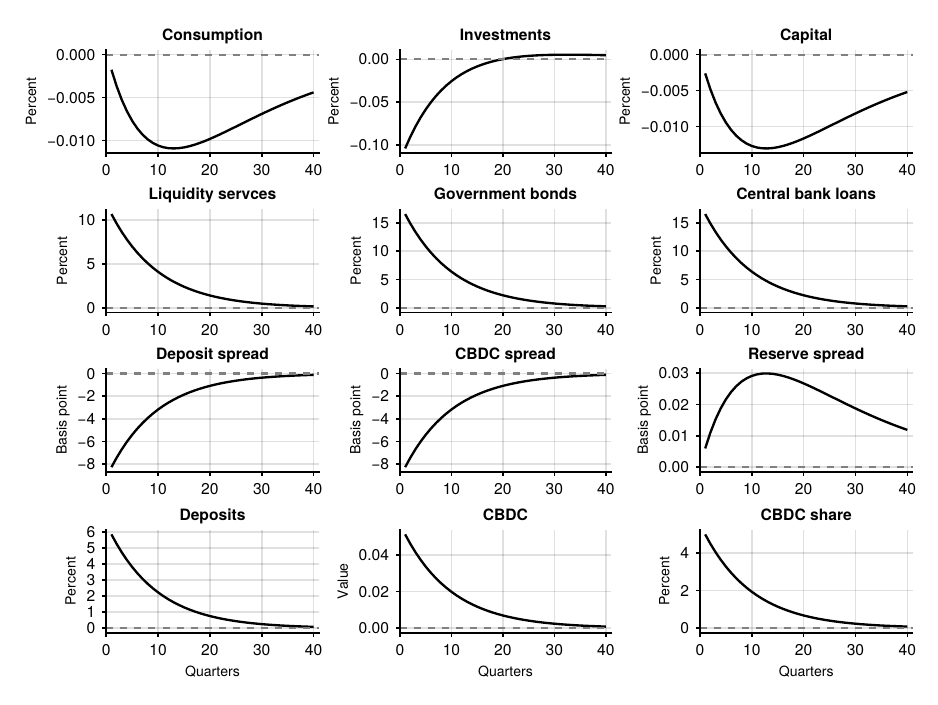}}
\end{minipage}
\vskip 1em
\begin{minipage}{\textwidth}
    \footnotesize
    \textit{Note}: This figure shows the impulse responses to a temporary increase in CBDC from zero to $5\%$ of steady-state output.
\end{minipage}
\label{fig:fig_temp_em}
\end{figure}

\begin{figure}[!htbp]
\caption{Impulse responses to a temporary increase in CBDC share from zero to 5\%}
\begin{minipage}{\textwidth}
    \centering
    \makebox[\textwidth]{\includegraphics[width=1\textwidth]{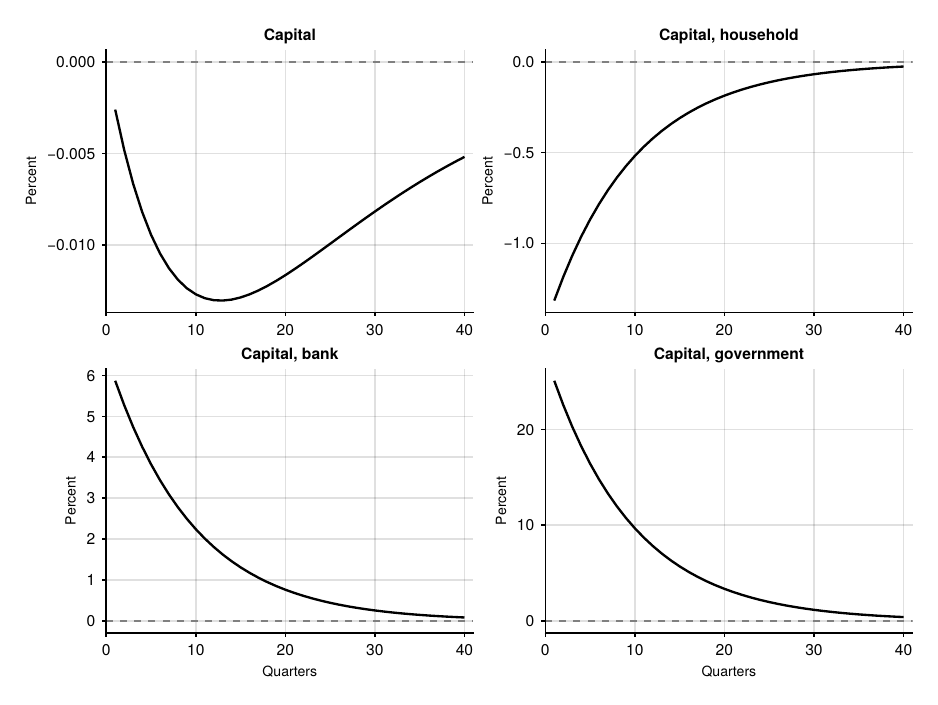}}\label{fig:temp_k_decomp}
\end{minipage}
\vskip 1em
\begin{minipage}{\textwidth}
    \footnotesize
    \textit{Note}: This figure shows the impulse responses to a temporary increase in CBDC from zero to $5\%$ of steady-state output.
\end{minipage}
\end{figure}

Figure \ref{fig:fig_temp_em} and \ref{fig:temp_k_decomp} show the impulse responses to a temporary increase in the CBDC share of steady-state output. Most variables react in similar ways to the baseline. The deposit spread falls, and bank intermediation (bank deposits and capital holdings) increases as the bank's balance sheet expands. Nevertheless, in this case, the increased credit intermediation of the bank and the government does not make up for the fall in household capital holdings, and consequently, capital and investments fall.

\newpage

\newpage

\subsection{Robustness} \label{sec:app:robustness}
\begin{figure}[!htbp]
\caption{Impulse responses to a near-permanent increase in CBDC share from zero to $5\%$ with lower relative benefit of CBDC ($\lambda=0.5$)}
\begin{minipage}{\textwidth}
    \centering
    \makebox[\textwidth]{\includegraphics[width=1\textwidth]{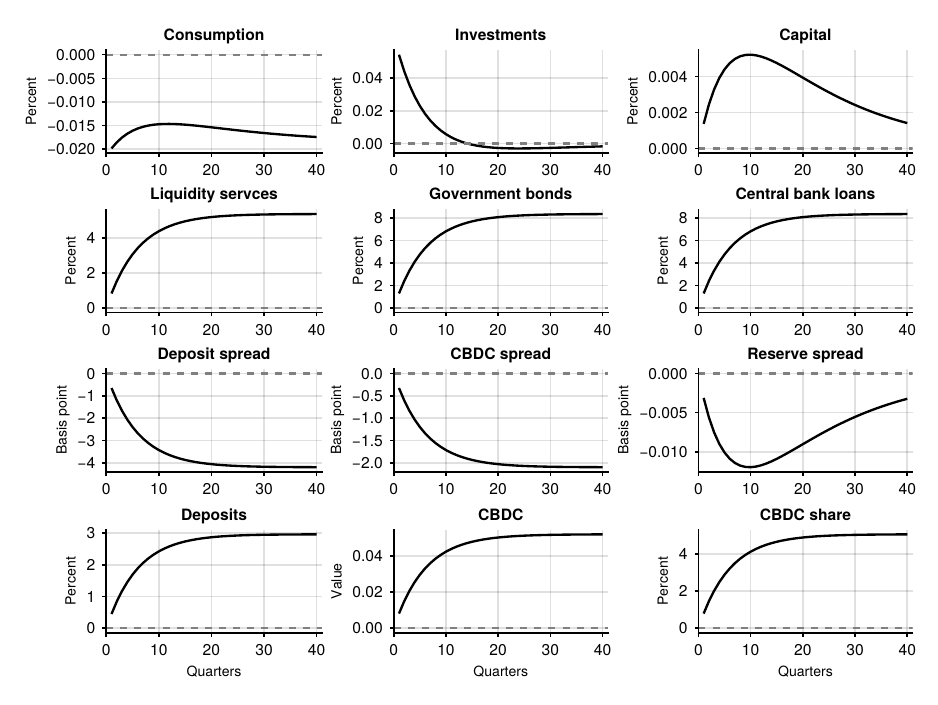}}
\end{minipage}
\vskip 1em
\begin{minipage}{\textwidth}
    \footnotesize
    \textit{Note}: This figure shows the impulse responses to a gradual and near-permanent increase in CBDC from zero to $5\%$ of steady-state output. The initial increase in the CBDC share is about 0.8 percentage points. The relative benefit of CBDC, $\lambda$, is set to $0.5$.
\end{minipage}
\label{fig:irf_lam05}
\end{figure}
\newpage

\begin{figure}[!htbp]
\caption{Impulse responses to a near-permanent increase in CBDC share from zero to $5\%$ with higher relative benefit of CBDC ($\lambda=1.5$)}
\begin{minipage}{\textwidth}
    \centering
    \makebox[\textwidth]{\includegraphics[width=1\textwidth]{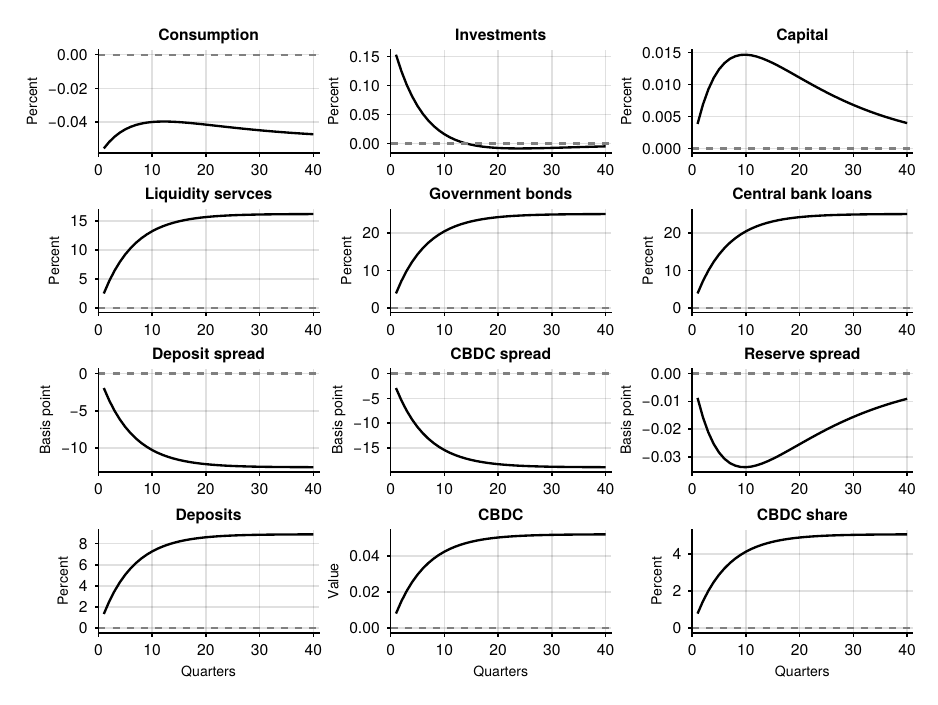}}
\end{minipage}
\vskip 1em
\begin{minipage}{\textwidth}
    \footnotesize
    \textit{Note}: This figure shows the impulse responses to a gradual and near-permanent increase in CBDC from zero to $5\%$ of steady-state output. The initial increase in the CBDC share is about 0.8 percentage points. The relative benefit of CBDC, $\lambda$, is set to $1.5$.
\end{minipage}
\label{fig:irf_lam15}
\end{figure}
\newpage

\begin{figure}[!htbp]
\caption{Impulse responses to a near-permanent increase in CBDC share from zero to $5\%$ with a lower haircut on bonds ($\theta_b=0.999$)}
\begin{minipage}{\textwidth}
    \centering
    \makebox[\textwidth]{\includegraphics[width=1\textwidth]{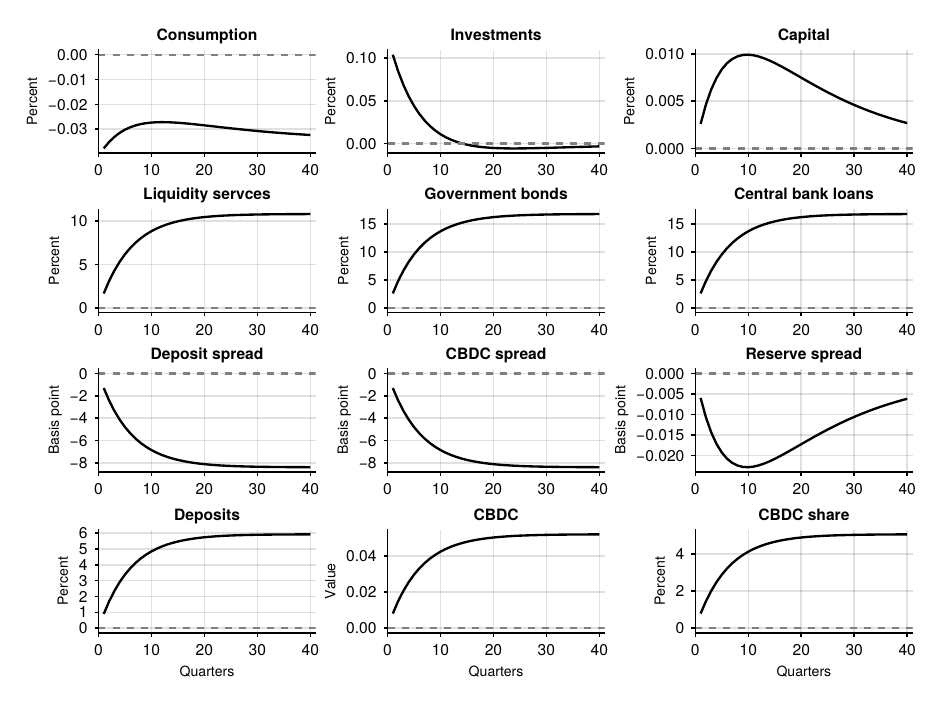}}
\end{minipage}
\vskip 1em
\begin{minipage}{\textwidth}
    \footnotesize
    \textit{Note}: This figure shows the impulse responses to a gradual and near-permanent increase in CBDC from zero to $5\%$ of steady-state output. The initial increase in the CBDC share is about 0.8 percentage points. The haircut on bonds, $\theta_b$, is set to $0.999$
\end{minipage}
\label{fig:irf_thb999}
\end{figure}
\newpage

\begin{figure}[!htbp]
\caption{Impulse responses to a near-permanent increase in CBDC share from zero to $5\%$ with a higher haircut on bonds ($\theta_b=0.985$)}
\begin{minipage}{\textwidth}
    \centering
    \makebox[\textwidth]{\includegraphics[width=1\textwidth]{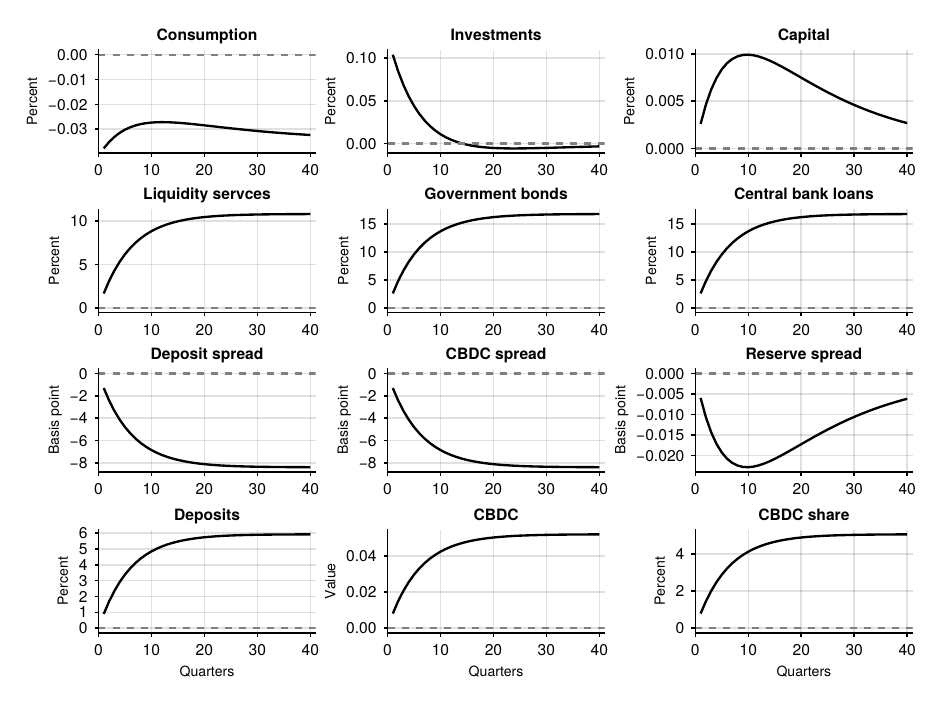}}
\end{minipage}
\vskip 1em
\begin{minipage}{\textwidth}
    \footnotesize
    \textit{Note}: This figure shows the impulse responses to a gradual and near-permanent increase in CBDC from zero to $5\%$ of steady-state output. The initial increase in the CBDC share is about 0.8 percentage points. The haircut on bonds, $\theta_b$, is set to $0.985$
\end{minipage}
\label{fig:irf_thb985}
\end{figure}

\end{document}